\documentclass[12pt]{article}
\usepackage[utf8]{inputenc}
\usepackage[english]{babel}
\usepackage{caption}
\usepackage{amsmath,amssymb,amsthm}
\usepackage{mathtools}
\usepackage{nicefrac}
\usepackage{hyperref}
\hypersetup{colorlinks = true, linkcolor=blue, citecolor=blue}
\usepackage{tikz-cd}
\usepackage{mathabx}
\usepackage{enumitem}
\usepackage{dsfont}
\usepackage[a4paper,width=150mm,top=25mm,bottom=25mm,bindingoffset=6mm]{geometry}
\usepackage{fancyhdr}
\usepackage{authblk}

\newcommand\smallO{
  \mathchoice
    {{\scriptstyle\mathcal{O}}}
    {{\scriptstyle\mathcal{O}}}
    {{\scriptscriptstyle\mathcal{O}}}
    {\scalebox{.7}{$\scriptscriptstyle\mathcal{O}$}}
  }

\theoremstyle{plain}
\newtheorem{theorem}{Theorem}[section]
\newtheorem{lemma}[theorem]{Lemma}
\newtheorem{corollary}[theorem]{Corollary}
\newtheorem{proposition}[theorem]{Proposition}

\theoremstyle{definition}
\newtheorem{definition}[theorem]{Definition}

\newtheorem*{remark*}{Remark}

\newcommand\R{\mathbb R} 
\newcommand\scpr{\boldsymbol{\cdot}} 
\newcommand\inner[1]{\langle #1 \rangle}
\newcommand\C{\mathbb C} 
\newcommand\restr[2]{{
  \left.\kern-\nulldelimiterspace 
  #1 
  \vphantom{\big|} 
  \right|_{#2} 
  }}

\newcommand\Z{\mathbb Z} 
\newcommand\N{\mathbb N} 

\newcommand\norm[1]{\lVert #1 \rVert} 

\newcommand\ra{\rightarrow}

\newcommand{\mres}{\mathbin{\vrule height 1.6ex depth 0pt width
0.13ex\vrule height 0.13ex depth 0pt width 1.3ex}}

\begin{document}
\title{Two-term asymptotics of the exchange energy of the electron gas on symmetric polytopes in the high-density limit 
}
\author{Thiago Carvalho Corso\thanks{Email: \url{thiago.carvalho@ma.tum.de}}}
\affil{Zentrum Mathematik, Technische Universit\"at M\"unchen, Germany}

\renewcommand\Affilfont{\itshape\small}

\maketitle
\begin{abstract} We derive a two-term asymptotic expansion for the exchange energy of the free electron gas on strictly tessellating polytopes and fundamental domains of lattices in the thermodynamic limit. This expansion comprises a bulk (volume-dependent) term, the celebrated Dirac exchange, and a novel surface correction stemming from a boundary layer and finite-size effects. Furthermore, we derive analogous two-term asymptotic expansions for semi-local density functionals. By matching the coefficients of these asymptotic expansions, we obtain an integral constraint for semi-local approximations of the exchange energy used in density functional theory.
\end{abstract}


\section{Introduction and main results}

We start with some notation and then present the main results of this paper. Let $\Omega \subset \R^n$ denote an open, bounded and connected subset with regular boundary. Then under either Dirichlet or Neumann boundary conditions (BCs), there exists a sequence $0\leq \lambda_1 \leq \lambda_2  \leq ... \ra \infty $ and an orthonormal basis (in $L^2(\Omega)$) of smooth functions $\{e_j\}_{j\in \N} \subset C^\infty(\Omega)$ such that 
\[ -\Delta e_j = \lambda_j^2 e_j  ,\]
where $\Delta$ is the Euclidean Laplacian \cite{taylor1996partial}. One can thus define the spectral function and its scaled diagonal up to $\lambda$ as
\begin{align} 
    S_\lambda(r,\tilde{r}) \coloneqq \sum_{\lambda_j \leq \lambda} e_j(r) \overline{e_j}(\tilde{r})  \quad \mbox{ and } \quad S_{s,\lambda}(r) \coloneqq \frac{1}{\lambda^n}S_\lambda\biggr(\frac{r}{\lambda},\frac{r}{\lambda}\biggr) . \label{eq:specfundef}
\end{align}
 In this paper, our goal is to derive two-term asymptotic expansions for the exchange energy with Riesz interaction,
\begin{align}  
    E_x(\lambda) = \int_{\Omega\times \Omega} \frac{|S_\lambda(r,\tilde{r})|^2}{|r-\tilde{r}|^s} \,\mathrm{d}r\,\mathrm{d}\tilde{r}  \quad \mbox{ with $0<s<n$,} \label{exchangespec} 
\end{align}
and for semi-local functionals
\begin{align} 
    F(\lambda) = \int_{ \Omega_\lambda} f\bigr(2S_{s,\lambda}(r), 2\nabla S_{s,\lambda}(r)\bigl) \,\mathrm{d}r , \label{semilocal}
\end{align}
in the limit $\lambda \ra \infty$. (The factor of $2$ inside $f$ comes from the spin of the electrons, see eq.~\eqref{eq:densityformula}.) Note that the class of semi-local functionals includes the important example of the counting function
\begin{align}
    N(\lambda) \coloneqq \#\{j :\lambda_j\leq \lambda\} = \int_{\Omega_\lambda} S_{s,\lambda}(r) \,\mathrm{d}r . \label{eq:countfuncdef}
\end{align}
From this example and the extensive literature on it (see \cite{Weyl12,ivrii2016100,ivrii2019microlocal,sogge2014hangzhou,safarov1997asymptotic,hormander2007analysis} and references therein), one sees that two-term asymptotics of this kind are often subtle and influenced by the regularity of the boundary and geometry of the domain. Even in the simple case of a connected domain with smooth boundary in $\R^n$, it is not known\footnote{Formula \eqref{weyl} is called the (two-term) Weyl law \cite{Weyl12} and is known to hold under some non-periodicity assumptions on the geodesic flow \cite{ivrii1980second,melrose1980weyl}. Such assumptions are conjectured to hold for general smooth domains but only proved (to the knowledge of the author) for special cases such as convex domains (see \cite{safarov1997asymptotic})} whether the following two-term asymptotic formula  holds:
\begin{align}  
    N(\lambda) = \begin{dcases} \frac{\omega_n}{(2\pi)^n} \lambda^n |\Omega| - \frac{\omega_{n-1}}{4(2\pi)^{n-1}} \lambda^{n-1} |\partial \Omega| + \smallO(\lambda^{n-1}) , &\mbox{ for Dirichlet BCs,} \\
    \frac{\omega_n}{(2\pi)^n} \lambda^n |\Omega| + \frac{\omega_{n-1}}{4(2\pi)^{n-1}} \lambda^{n-1} |\partial \Omega| + \smallO(\lambda^{n-1}) , &\mbox{ for Neumann BCs,} \end{dcases}
\label{weyl} 
\end{align}
where $\omega_n$ is the volume of the unit ball on $\R^n$, $|\Omega|$ is the volume of $\Omega$, and $|\partial \Omega|$ is the area of the boundary. Therefore, we restrict ourselves here to two types of domains where such asymptotics can be obtained: (i) the set of strictly tesselating polytopes $\Omega \subset \R^n$ (see Definition \ref{def:kaleidoscopic}), and (ii) fundamental domains of lattices $\Gamma \subset \R^n$ with the periodic Laplacian.

For such domains, the main theorems of this paper read as follows.
\begin{theorem}[Asymptotics of exchange energy]\label{exchangethm}
Let $\Omega \subset \R^n$ be a strictly tessellating polytope (see Definition \ref{def:kaleidoscopic}) or the fundamental domain of a lattice. Let $E_x(\lambda)$ be the exchange energy defined in \eqref{exchangespec} and suppose that $n\geq 2$ and $\frac{n-1}{2}-\frac{n-1}{n+1}<s<n$. Then, for any $\epsilon>0$ we have
\begin{align} E_x(\lambda) = c_{x,1}(n,s) \lambda^{n+s} |\Omega| + \bigr(c_{FS}(n,s) + c_{BL}(n,s)\bigr) \lambda^{n-1+s} |\partial \Omega| + \mathcal{O}(\lambda^{r(n,s)+\epsilon}) , \label{asympexchange} \end{align}
where 
\begin{align*}
    r(n,s) = \begin{cases} \max\{s,7/6,1+s/6\} &\mbox{ for $n = 2$,}\\
    \max\{n-2+s,(3n-2)/2-(n-1)/(n+1)\} &\mbox{otherwise.}\end{cases}
\end{align*}
The leading exchange, the finite size, and the boundary layer constants are given by
\begin{align*}
    &c_{x,1}(n,s) = \frac{\omega_n^2}{(2\pi)^{2n}}\int_{\R^n} \frac{h_n(|z|)^2}{|z|^s} \,\mathrm{d}z,  \quad \quad  c_{FS}(n,s) = -\frac{\omega_n^2}{(2\pi)^{2n}}\int_{\R^n} \frac{h_n(|z|)^2|z_n|}{2|z|^s}\,\mathrm{d}z, \\
   &c_{BL}^{\mathrm{Per}}(n,s) = 0, \\
   &c_{BL}^{\mathrm{Dir}}(n,s) = \frac{ \omega_n^2}{(2\pi)^{2n}}\int_{\R^n} \int_{|z_n|}^\infty h_n(|(\pi_n z, w_n)|)\frac{h_n(|(\pi_n z, w_n)|)  - 2 h_n(|z|)}{2|z|^s} \,\mathrm{d}w_n\,\mathrm{d}z, \\
   &c_{BL}^{\mathrm{Neu}}(n,s) =\frac{\omega_n^2}{(2\pi)^{2n}}\int_{\R^n} \int_{|z_n|}^\infty h_n(|(\pi_n z, w_n)|)\frac{h_n(|(\pi_n z, w_n)|)  + 2 h_n(|z|)}{2|z|^s} \,\mathrm{d}w_n\,\mathrm{d}z,
\end{align*}
where $\omega_n = |B_1|$ is the volume of the unit ball in $\R^n$, $h_n(|r|) = \widehat{\chi_{B_1}}(r)/\omega_n$ is the normalized Fourier transform of the characteristic function of $B_1$, $\pi_n$ is the projection on the first $n-1$ coordinates, and the superscripts indicate the boundary conditions.
\end{theorem}

\begin{theorem}[Asymptotics of semi-local functionals]\label{semilocalthm} Let $\Omega \subset \R^n$ be a strictly tessellating polytope. Suppose that $f\in C^1\bigr((0,\infty)\times \R^n\bigr)\cap L^\infty_{\textnormal{loc}}([0,\infty)\times \R^n)$. Then, for $F(\lambda)$ given by $f$ via \eqref{semilocal}, we have
\begin{align} 
    F(\lambda) = f(\nu_0) \lambda^n |\Omega| + c(f,\Omega) \lambda^{n-1} + \smallO(\lambda^{n-1}), \label{eq:semilocalasymp}
\end{align}
where
\begin{align*}
    c(f,\Omega) = \begin{dcases} \int_{\partial \Omega} \biggr(\int_0^\infty f(\nu_0 - \nu_1(\tau,r^\prime))-f(\nu_0) \,\mathrm{d} \tau\biggl)\,\mathrm{d}\mathcal{H}^{n-1}(r^\prime) , &\mbox{for Dirichlet BCs,} \\
    \int_{\partial \Omega} \biggr(\int_0^\infty f(\nu_0 + \nu_1(\tau,r^\prime))-f(\nu_0) \,\mathrm{d} \tau \biggl)\,\mathrm{d} \mathcal{H}^{n-1}(r^\prime) , &\mbox{for Neumann BCs,}
    \end{dcases}
\end{align*}
and
\begin{align*}
    \nu_0  \coloneqq \frac{2\omega_n}{(2\pi)^n}(1,0) \in \R\times \R^{n}, \quad \nu_1(\tau,r^\prime) \coloneqq \frac{ 2\omega_n}{(2\pi)^n}\bigr(h_n(2 \tau),2\dot{h_n}(2 \tau)n(r^\prime)\bigr),
\end{align*}
where $n(r^\prime)$ is the inwards pointing unit normal to $\partial \Omega$ at $r^\prime$, $\mathcal{H}^{n-1}$ is the $(n-1)$-dimensional Hausdorff measure, and $\omega_n$ and $h_n$ are the same from Theorem~\ref{exchangethm}.
\end{theorem}

\begin{figure}[ht!]
    \centering
    \includegraphics[scale=0.3]{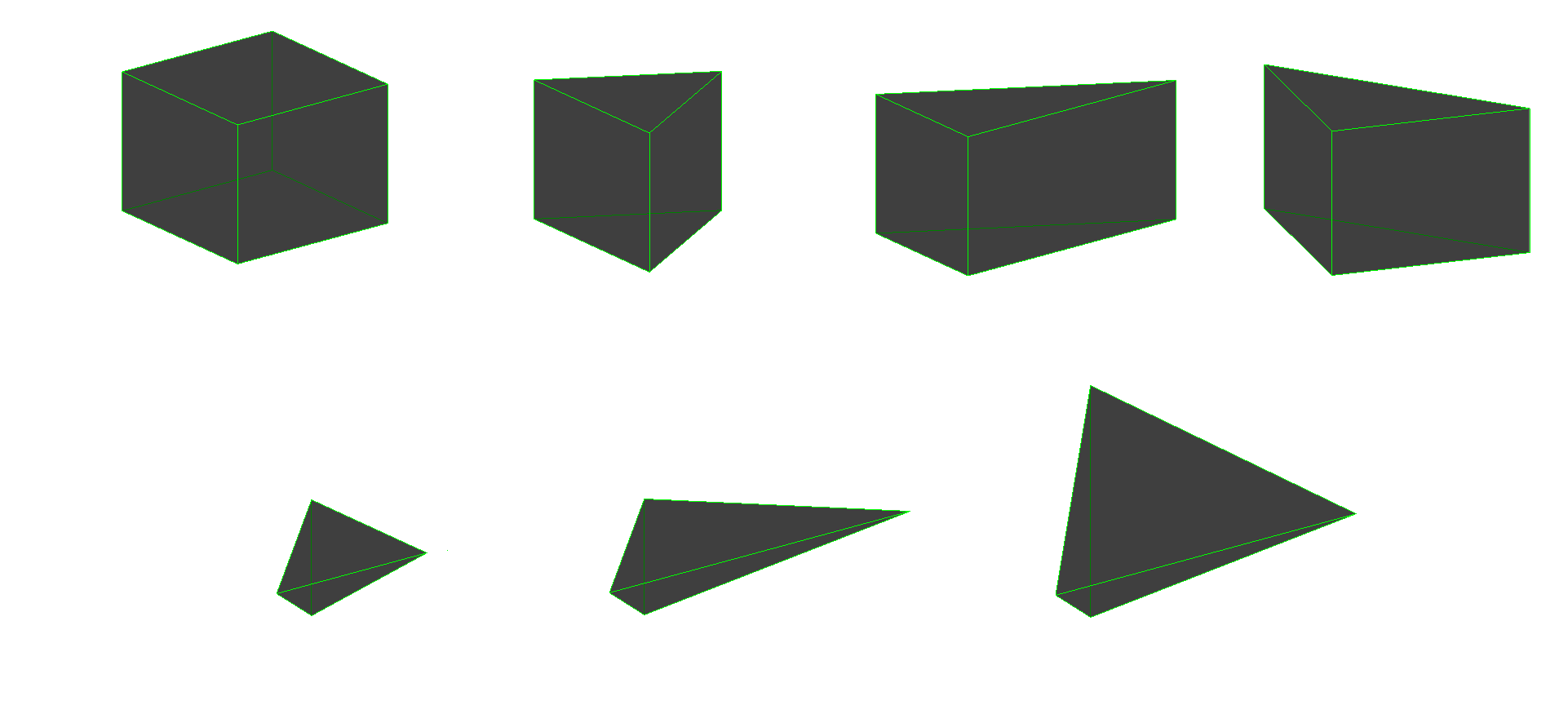}
    \caption{Kaleidoscopic polytopes in $\R^3$. From left upper corner: rectangular parallelepiped, equilateral prism, 30-60-90 prism, isosceles (45-45-90) prism, quadrirectangular tetrahedron, trirectangular tetrahedron, and tetragonal disphenoid.}
    \label{figure:kaleidoscopicpolytopes}
\end{figure}

\begin{remark*}[On the constants] Due to radial symmetry of the interaction $1/|r|^s$, the coefficients $c_{x,1}(n,s)$, $c_{FS}(n,s)$ and $c_{BL}(n,s)$ can be computed by numerically evaluating a 1D, 2D and 3D integral respectively. For the Coulomb interaction in 3D, the constants can be analytically computed  (see \cite[Lemma 5.2]{corso2023nextorder}), and their values are
\begin{align}
	c_{x,1}(3,1) = \frac{1}{4\pi^3}, \quad c_{FS}(3,1) = -\frac{1}{24 \pi^2},\quad\mbox{and}\quad c_{BL}^{Dir}(3,1) = -\frac{\log 2}{12 \pi^2}. \label{eq:constants}
\end{align}
\end{remark*}
\begin{remark*}[Periodic case] The asymptotics of semi-local functionals for the periodic case is trivial and has no boundary corrections because $S_{s,\lambda}(r) = 
 N(\lambda)/\lambda^n =  
 \omega_n/(2\pi)^n + \mathcal{O}(\lambda^{-1-\frac{n-1}{n+1}})$ in this case. Moreover, the seemingly unphysical boundary correction for the exchange energy in the periodic case comes from the fact that the interaction potential is not periodic.
\end{remark*}

\begin{remark*} Theorems~\ref{exchangethm} and ~\ref{semilocalthm} can be extended in the following directions:\begin{enumerate}[label=(\roman*)]
\item {\textrm (Interactions)} Theorem~\ref{exchangethm} can be directly extended to non-radial interactions $w$ satisfying $c/|r|^s \leq w(r) \leq C/|r|^s$ for some positive constants $c,C>0$ (e.g. positively homogeneous interactions).
\item {\textrm (Mixed boundary conditions)} Both theorems can be extended to the situation where Dirichlet and Neumann BCs are imposed on different faces of $\Omega$. This can be done by modifying the factor $\det \sigma$ appearing in the generalized Poisson summation (cf. Corollary~\ref{cor:generalizedpoisson}) to account only for the reflections over Dirichlet faces. 
\item {\textrm (Smooth domains)} Under the stronger condition that $f \in L^\infty_{\mathrm{loc}}([0,\infty)\times \R^3) \cap C^2((0,\infty)\times \R^3)$, the asymptotics of $F(\lambda)$ can be extended to smooth domains for which the two-term Weyl law \eqref{weyl} holds. More precisely, this can be achieved by using the gradient estimates in \cite{shi2013gradient} to extend Theorem 17.5.10 in \cite[Chapter XVII, pp.52]{hormander2007analysis} to first-order derivatives and following the same steps from the proof of Theorem~\ref{semilocalthm} (see Section~\ref{sec:asympexchangefunctionals}).
\item {\textrm (Beyond gradient approximations)} By using the $C^\infty$-convergence of the spectral function (cf. Theorem~\ref{offdiagonalthm}), we can also obtain two-term expansions for semi-local density functionals that depend on higher-order derivatives of the spectral function, such as the meta-generalized gradient approximations (mGGAs) used in density functional theory (DFT) \cite{TaoPerdew2003,Sun2016}.
\end{enumerate}
\end{remark*}

\textit{Applications to quantum mechanics and DFT.} Let us now briefly explain how the quantities $E_x(\lambda)$ and $F(\lambda)$ in the limit $\lambda \ra \infty$ are related to the exchange energy of the electron gas in the high-density limit. The ground state energy of the \emph{interacting} $N$-electron gas (under Dirichlet boundary conditions\footnote{From the physics point of view, the Dirichlet boundary conditions arise naturally when the external potential is set to $+\infty$ outside $\Omega_L$ and zero inside.}) confined to a domain $\Omega_L = \{ r/L \in \Omega\} \subset \R^3$ is defined as
\begin{multline*}
    \mathcal{E}_{N,L} = \inf_{\substack{\Psi \in \mathcal{H}_N(\Omega_L) \\ \norm{\Psi}_{L^2(\Omega_L)} =1 }} \underbrace{\frac{1}{2}\int_{(\Omega_L\times \mathcal{S})^N}|\nabla \Psi(x_1,...,x_N)|^2 \,\mathrm{d}x_1...\,\mathrm{d}x_N}_{=:T[\Psi]} \\ 
    + \underbrace{\int_{(\Omega_L \times \mathcal{S})^N}\sum_{1\leq i < j \leq N} \frac{|\Psi(x_1,...,x_N)|^2}{|r_i-r_j|}\,\mathrm{d}x_1...\,\mathrm{d}x_N}_{ =: V_{\textrm ee}[\Psi]},
\end{multline*}
where $\mathcal{S} = \{0,1\}$ denote the spin states, $\mathcal{H}_N(\Omega_L) \subset L^2((\Omega_L \times \mathcal{S})^N)$ is the set of anti-symmetric (with respect to the spin variables $x_\ell = (r_\ell, s_\ell) \in \Omega_L \times \mathcal{S}$) Sobolev functions satisfying Dirichlet boundary conditions, and $\int_{\Omega_L \times \mathcal{S}} \mathrm{d}x = \sum_{s \in \mathcal{S}} \int_{\Omega_L} \mathrm{d}r$ where $\mathrm{d}r$ is the standard Lebesgue measure in $\R^3$. From a simple scaling argument, we find
\begin{align*}
\mathcal{E}_{N,L} = \inf_{\substack{\Psi \in \mathcal{H}_N(\Omega) \\ \norm{\Psi}_{L^2(\Omega)} = 1}} L^{-2} T[\Psi] + L^{-1} V_{ee}[\Psi].
\end{align*}
Moreover, under the closed shell condition $N = 2N(\lambda)$ for some $\lambda>0$, the ground state of the free $N$-electron gas (FEG) on $\Omega$, i.e., the minimizer $\Psi_\lambda$ of $T$ over normalized functions in $\mathcal{H}_N(\Omega)$ is the unique anti-symmetric $N$-fold tensor product (or Slater determinant) of the orbital functions
\begin{align} \phi_{\ell}(r,s) = e_{\lfloor \ell/2 \rfloor}(r) \chi_{\ell  - 2\lfloor \ell/2  \rfloor}(s) \in L^2(\Omega \times \mathcal{S})\quad \ell \leq 2N(\lambda), \label{eq:orbital}\end{align}
where $\lfloor \ell/2 \rfloor$ is the greatest integer smaller than or equal to $\ell/2$ (the floor function) and $\chi_j(s) = 1$ for $s = j$ and zero otherwise (see, e.g. \cite[Section 2]{Friesecke1997} for a proof). Consequently, standard perturbation theory and straightforward calculation yields
\begin{align} 
\lim_{L \ra 0} \frac{L^2 \mathcal{E}_{2N(\lambda),L} 
 -T[\Psi_\lambda]}{L}  - \underbrace{\frac12 \int_{\Omega \times \Omega} \frac{\rho_{\lambda}(r) \rho_{\lambda}(r')}{|r-r'|}\,\mathrm{d}r \,\mathrm{d} r'}_{=:J[\rho_{\lambda}]} =  -E_x(\lambda), \label{eq:exchangehighdensity}
\end{align}
where 
\begin{align} \rho_\lambda(r) = N \sum_{s_1,..,s_N \in \mathcal{S}} \int_{\Omega^N} |\Psi_\lambda(r,s_1,r_2,s_2...,r_N,s_N)|^2\,\mathrm{d}r_2... \,\mathrm{d}r_N = 2S_\lambda(r,r) \label{eq:densityformula}
\end{align}
is the single-particle density of the FEG $\Psi_\lambda$, $J$ is the Hartree (or direct) term, and $E_x(\lambda)$ is the exchange energy defined in \eqref{exchangespec}.  
Similarly, we can relate the function $F(\lambda)$ defined in \eqref{semilocal} with generalized gradient approximations (GGA) for the exchange energy. In the physics literature \cite{Becke1988,PBE1996}, exchange GGAs are expressed as
\begin{align}
    E^{\textrm GGA}_x[\rho] = - \int_{\Omega} c_x \rho(r)^{\frac43} F_x\bigr(s(r)\bigr) \,\mathrm{d} r, \label{eq:GGAdef}
\end{align}
where $c_x = (3/\pi)^{\frac13}3/4$ is the Dirac constant, $s(r) = |\nabla \rho(r)|/\rho(r)^{\frac43}$ is the dimensionless reduced gradient and the function $F_x : [0,\infty) \rightarrow \R$ is called the enhancement factor and satisfy $F_x(0) = 1$\footnote{In particular, $E^{\textrm GGA}_x[\rho]$ reduces to the local density approximation (LDA) of the exchange energy given by the celebrated Dirac-Bloch exchange formula \cite{Bloch1929,dirac_1930}.}. Thus from another scaling argument, we find that
\begin{align}
    E_x^{\textrm GGA}[\rho_{\lambda}] = \lambda F(\lambda), \label{eq:GGAsemilocalrelation}
\end{align}
where $F$ is defined by \eqref{semilocal} with $f(2\rho, 2\nabla\rho) = -c_x \rho^{\frac43} F_x(|\nabla \rho|/\rho^{\frac43})$. We can now use Theorems~\ref{exchangethm} and \ref{semilocalthm} ,eqs.~\eqref{eq:exchangehighdensity} and \eqref{eq:GGAsemilocalrelation}, and the values in \eqref{eq:constants} to obtain
\begin{corollary}[Integral constraint for GGAs]\label{cor:ggaconstraint} Let $\Omega\subset \R^3$ be a strictly tessellating polyhedron, $\rho_\lambda$ be the single-particle density of the unique minimizer of $T$ on $\mathcal{H}_{2N(\lambda)}(\Omega)$, and $f(2a,2b) = -c_x a^{\frac 43} F_x(|b|/a^{\frac43}) \in C^1((0,\infty)\times \R^3)\cap L^\infty_{\textrm loc}([0,\infty)\times \R^3)$ with $F_x(0) =1$. Then we have
\begin{align*}
    \lim_{L \ra 0} \frac{L^2\mathcal{E}_{2N(\lambda),L} -T[\Psi_\lambda]}{L} - J[\rho_\lambda] = E^{\textrm GGA}_x[\rho_\lambda] +\smallO(\lambda^2)
\end{align*}
if and only if
\begin{align}
    \frac{1}{2(3\pi^2)^{\frac13}}\int_0^\infty\biggr[1- \bigr(1-h_3(\tau)\bigr)^{\frac43}F_x\biggr(2(3\pi^2)^{\frac13}\frac{|\dot{h_3}(\tau)|}{(1-h_3(\tau))^{\frac43}}\biggr) \biggr]\,\mathrm{d} \tau = \frac{1+\log 2}{8 c_x},  \label{eq:GGAconstraint}
\end{align}
where $c_x = (3/\pi)^{\frac13}3/4$ is the Dirac constant and $h_3(\tau) = 3(\sin \tau - \tau \cos \tau)/\tau^3$.
\end{corollary}

\begin{remark*}[Kinetic energy approximations] Two-term asymptotics of the kinetic energy $T[\Psi_\lambda]$, which is simply the sum of the eigenvalues of the Laplacian up to $\lambda$, are well-known \cite{ivrii2016100} (even under weak assumptions on $\partial \Omega$ \cite{Frank_2011,FrankLarson2020}). Therefore, Theorem~\ref{semilocalthm} can also be used to obtain an integral constraint on semi-local approximations of the kinetic energy, which play a central role in orbital-free DFT \cite{WangOFDFT2013}.
\end{remark*}

\textit{Proof strategy.} The underlying strategy in the proofs of Theorems \ref{exchangethm} and \ref{semilocalthm} is the same and consists of two main steps: (i) we obtain precise asymptotics for the spectral function, including the behaviour close to the boundary, and (ii) we perform a careful analysis of the interior and boundary terms. 

The first step is done via the wave equation (or kernel) method. To construct the exact wave kernel for all times, we use the symmetries of the domain $\Omega$. At this step, the reflection (respectively, translation) symmetry of the strictly tessellating polytopes (respectively, fundamental domains of lattices) plays a central role and is the main reason for our restriction to such domains. With the exact wave kernel at hand, we follow the approach in \cite[Chapter 3]{sogge2014hangzhou} to obtain the continuum limit of the spectral function with explicit uniform estimates. Such estimates include derivatives and are not restricted to the diagonal; they can be stated as follows.
\begin{theorem}[Asymptotics of the spectral function] \label{offdiagonalthm} Let $\Omega \subset \R^n$ be a strictly tessellating polytope or a fundamental domain of a lattice. Then for any $\alpha,\beta \in \N_0^n$, there exists a constant $C= C(\Omega,\alpha,\beta)>0$ such that 
\begin{align} 
    \bigr|\partial^\alpha_r\partial^\beta_{r'} S_\lambda(r,r')-\partial^\alpha_r \partial^\beta_{r'} S_\lambda^{\textnormal{ctm}}(r,r') \bigr|\leq C\left( 1+\lambda^{n-1-\frac{n-1}{n+1} + |\alpha|+|\beta|}\right) ,\label{specasympdir}
\end{align}
where
\begin{align}
    S_\lambda^{\textnormal{ctm}}(r,r') = \begin{dcases}  \frac{\omega_n}{(2\pi)^n} \lambda^n \sum_{v \in \mathcal{T}_{\Omega}^{nb}} h_n(\lambda |r-r'+v|) &\mbox{for periodic BCS,} \\
    \frac{\omega_n}{(2\pi)^n} \lambda^n\sum_{\sigma \in \mathcal{R}_{\Omega}^{nb}} \det \sigma h_n(\lambda |r-\sigma r'|)  &\mbox{for Dirichlet BCs,}\\
    \frac{\omega_n}{(2\pi)^n} \lambda^n\sum_{\sigma \in \mathcal{R}_{\Omega}^{nb}} h_n(\lambda |r-\sigma r'|)\bigr) &\mbox{for Neumann BCs,} \end{dcases}
\end{align}
where $\omega_n$ and $h_n$ are the same from Theorem~\ref{exchangethm}, $\mathcal{T}_\Omega^{nb}$ and $\mathcal{R}_\Omega^{nb}$ are, respectively, the sets of neighbouring translations and reflections of $\Omega$, and $\det \sigma$ is the determinant of the linear part of $\sigma$. (See \eqref{eq:nbreflectiondef} and the preceding discussion for the proper definitions.)
\end{theorem}

Estimate~\eqref{specasympdir} is enough to justify the use of the continuum approximation $S_\lambda^{\textnormal{ctm}}$ for the asymptotics of $F(\lambda)$; this follows by using the Lipschitz regularity of the function $f$ in the integrand of $F(\lambda)$, and a cut-off away from the boundary to avoid the points where $\rho =0$ and $f$ is no longer Lipschitz (see Section~\ref{sec:asympexchangefunctionals}). 

On the other hand, the above estimates are not enough to justify the use of the continuum approximation for the exchange energy. Roughly speaking, this is because the exchange energy is given by integration against the square of the spectral function. Therefore, the error estimate in \eqref{specasympdir} yields an error proportional to $(\lambda^{\frac32})^2 = \lambda^3$ (in the 3D Coulomb case) between the exchange energy of the spectral function and its continuum version, which is precisely the order of the second term in Theorem \ref{exchangethm}. In \cite{corso2023nextorder}, where the case $\Omega = [0,1]^3$ was studied, the authors overcame this problem by using the theory of exponential sums to improve the remainder in \eqref{specasympdir} from $\lambda^{\frac32}$ to $\lambda^{\frac32 - \frac{1}{46}+\epsilon}$. This was possible because explicit eigenfunction formulae are available in the rectangular box. In this paper, however, we aim to derive such asymptotics without explicit expressions for the eigenfunctions. Inspired by the work in \cite{schmidt2011localized}, we realized that interpolating the $L^\infty$ estimates from Theorem \ref{offdiagonalthm} with $L^2$ estimates is a much more efficient approach for two reasons: first, the $L^2$ estimates can be obtained by slightly modifying the proof of the $L^\infty$ estimates; and second, they lead to a significant improvement in the remainder of the asymptotic expansion of the exchange energy. Our main estimate in the $L^2$ setting is the following.
\begin{theorem}[$L^2$ estimate of spectral function]\label{L2thm}  Let $\Omega \subset \R^n$ be a strictly tessellating polytope or a fundamental domain of a lattice. Then, there exists $C = C(n,\Omega) >0$ such that
\begin{align} 
    \norm{S_\lambda - S^{\textnormal{ctm}}_\lambda}_{L^2(\Omega\times \Omega)} \leq C(1+\lambda^{\frac{n-1}{2}}), \label{L2conv} \end{align}
where $S^{\textnormal{ctm}}_\lambda$ is the same from Theorem \ref{offdiagonalthm}.
\end{theorem}
By combining Theorems \ref{offdiagonalthm} and \ref{L2thm}, we can justify the use of the continuum spectral function to evaluate the exchange energy. The asymptotic expansion for $E_x(\lambda)$ then follows from geometric considerations and a careful analysis of the boundary and interior terms (see Section~\ref{sec:asympexchangefunctionals}).

\textit{Related works.} The literature on asymptotics of the spectral function of the Laplacian is vast (see \cite{ivrii2016100} for a recent review). In the interest of time, we shall only mention the works most related to Theorems~\ref{offdiagonalthm} and \ref{L2thm}. First, the diagonal version of Theorem~\ref{offdiagonalthm} with $\alpha =\beta =0$ for the Torus is well-known and can be found, e.g., in \cite{sogge2014hangzhou,sogge2017fourier}. The extension to higher-order derivatives and for general manifolds is also known; see, e.g., \cite{Canzani2018Scaling}, or \cite[Section 17]{hormander2007analysis}. In these works, however, the remainder is of order $\lambda^{n-1}$ (which is known to be sharp in some cases) and degenerates close to the boundary. Improvements over this sharp remainder are associated with dynamical properties of the geodesic flow  \cite{Duistermaat75,Berard1977,ivrii1980second,safarov1997asymptotic}, which makes the extension of the two-term asymptotics derived here to general manifolds a challenging problem. Concerning the $L^2$-estimates, similar results for manifolds without boundary can be found in \cite{Lapointe2009}. Finally, let us mention the work by B\'erard \cite{Berard1980-xg}, which appears to contain very similar results to the ones proved in Theorem~\ref{offdiagonalthm}. Unfortunately, we could not find an English version of \cite{Berard1980-xg} to properly compare the methods used there with the ones here.   

The leading order asymptotics of the exact exchange energy and of local density approximations of the exchange energy of the FEG were studied in \cite{Friesecke1997}, for the rectangular box in $\R^3$, and in \cite{schmidt2011localized} for general domains $\Omega \subset \R^3$. Similar leading asymptotics of the exchange energy also appears in \cite{graf1994correlation}, where the electron gas on a constant neutralizing background is studied by first taking the thermodynamic limit and then the high-density limit. 

The next-order asymptotics for the exchange energy and semi-local approximations was derived for the first time in \cite{corso2023nextorder} in the rectangular box in $\R^3$. We remark, however, that the thermodynamic limit considered in \cite{corso2023nextorder} is slightly different from the one considered here. While in \cite{corso2023nextorder}, the authors fix the average density and consider the limit $N \ra \infty$; here, we use the Fermi momentum $\lambda$ as the asymptotic parameter. From the mathematical perspective, the asymptotics presented here are more natural because the Fermi momentum correction appearing in \cite[Lemma 3.2]{corso2023nextorder} is no longer necessary here. Nonetheless, we emphasize that the integral constraint for semi-local approximations obtained in Corollary~\ref{cor:ggaconstraint} is the same as the one proposed in \cite{corso2023nextorder}.

\textit{Structure of the paper.} In Section~\ref{sec:wavekernelpolytope}, we construct the exact wave kernel and derive a generalized Poisson summation formula on strictly tessellating polytopes. We then use this Poisson summation formula to prove Theorems~\ref{offdiagonalthm} and \ref{L2thm} in Section~\ref{sec:spectralfunction}. The proof of Theorems~\ref{exchangethm} and ~\ref{semilocalthm} are given in Section~\ref{sec:exchangepoly}. In Appendix~\ref{sec:tesselatingpolytopes}, we show that the definition of strictly tessellating polytopes presented here is equivalent to \cite[Definition 2]{rowlett2021crystallographic}. In Appendix~\ref{sec:wavekernel}, we collect some well-known facts about the wave equation that are used throughout the proofs. 

\subsection*{Notation}

Throughout this paper, $\Omega \subset \R^n$ denotes a bounded, connected, and open subset of $\R^n$, where $n\geq 2$. The re-scaled version of $\Omega$ by a factor $L>0$ is denoted by $\Omega_L = \{r\in \R^n: r/L \in \Omega\}$. The characteristic function of a set $\Omega \subset \R^n$ is denoted by $\chi_\Omega$. The unit ball in $\R^n$ is denoted by $B_1$. For the Fourier transform of a function $f: \R^n \rightarrow \C$, we use the convention
\begin{align*}
	\widehat{f}(k) = \int_{\R^n} f(r) e^{-i k \scpr r} \mathrm{d}r,
\end{align*}
where $k \scpr r = \sum_{j=1}^n k_j r_j$ is the standard scalar product in $\R^n$. The Schwartz space of test functions and tempered distribution in $\R^n$ are denoted, respectively, by $S(\R^n)$ and $S'(\R^n)$. We use the standard big-O and small-O notation. More precisely, for functions $f: [0,\infty) \rightarrow \R$ and $g: [0,\infty) \rightarrow \R$ we say that $f = \mathcal{O}(g)$ respectively $f = \smallO(g)$ provided that
\begin{align*}
	\limsup_{\lambda \ra \infty} \frac{|f(\lambda)|}{|g(\lambda)|} < \infty \quad\mbox{respectively}\quad \limsup_{\lambda \ra \infty} \frac{|f(\lambda)|}{|g(\lambda)|} = 0.
\end{align*}
We also use the notation $f \lesssim g$ to indicate the existence of an unimportant constant $C>0$ such that $|f(\lambda)| \leq C |g(\lambda)|$ for all values of $\lambda$ large enough. In addition, if $f$ or $g$ depends on additional parameters (e.g. $\epsilon$), we indicate the dependence of the constant $C$ on this parameter by using the notation $f \lesssim_{\epsilon} g$.

\section{Wave kernel on symmetric polytopes}\label{sec:wavekernelpolytope}

We now turn to the construction of the wave kernel on strictly tessellating polytopes and fundamental domains of lattices. The key idea is to exploit the symmetries of the reflection/translation group associated to such polytopes. Let us start by introducing some notation and the proper definitions.

\begin{figure}[ht!]
    \centering
    \includegraphics[scale=0.38]{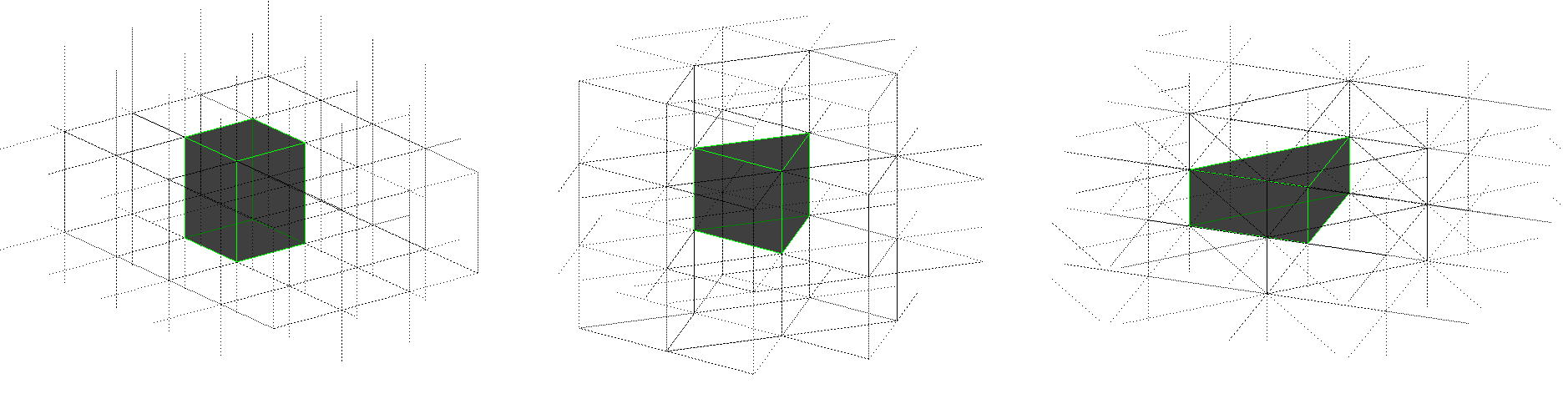}
    \caption{Example of strict tessellations of the space by some solids of Figure~\ref{figure:kaleidoscopicpolytopes}.}
    \label{figure:tessellates}
\end{figure}

Let $\Omega$ be a polytope in $\R^n$. We denote by $\{F_1, ..., F_m\}$ the set of boundary faces of $\Omega$, and by $\{\sigma_1, ..., \sigma_m\}$ the corresponding set of reflections over the faces of $\Omega$. 
The group of reflections, $\mathcal{R}_\Omega$, is then defined as the group generated by the reflections  $\{\sigma_{\ell}\}_{1\leq \ell \leq m}$ through composition, i.e., 
\begin{align}
    \mathcal{R}_\Omega = \{  \tau : \R^n \rightarrow \R^n : \tau = \sigma_{j_1} \circ ... \circ \sigma_{j_M}, \mbox{ where $j_k \in \{1,..., m\}$} \} . \label{eq:reflectiongroupdef}
\end{align}
For any $\sigma \in \mathcal{R}_\Omega$, we denote the determinant of the linear part of $\sigma$ by $\det \sigma$. Note that $\det \sigma \in \{1, -1\}$ for any $\sigma \in \mathcal{R}_\Omega$. 
The set of strictly tessellating polytopes can then be defined as follows.
\begin{definition}[Strictly tesselating polytopes]\label{def:kaleidoscopic} We say that an open polytope $\Omega \subset \R^n$ strictly tessellates $\R^n$ if for any $\sigma, \tau \in \mathcal{R}_\Omega$ with $\sigma \neq \tau$, the reflected polytopes $\sigma(\Omega)$ and $\tau(\Omega)$ do not intersect. In mathematical terms, $\Omega$ is strictly tessellating if and only if the following holds:
\begin{align}
    \sigma(\Omega) \cap \tau(\Omega) \neq \emptyset \iff \tau = \sigma. \label{eq:tessellate}
\end{align}
(See Figure ~\ref{figure:tessellates}.)
\end{definition}
\begin{remark*} The term strictly tessellates is adopted from \cite{rowlett2021crystallographic}. Note, however, that the definition given here is different from the one in \cite[Definition 2]{rowlett2021crystallographic}. The definition above is more convenient for our purposes, as can be seen from the proof of Lemma~\ref{lem:wavekernelkaleidoscopic} below. That both definitions are equivalent is shown in Section~\ref{sec:tesselatingpolytopes} 
\end{remark*}

Similarly, we can define the fundamental domain of a lattice $\Gamma$ as follows.
\begin{definition}[Fundamental domains] We say that an open polytope $\Omega \subset \R^n$ is the fundamental domain of a lattice $\Gamma = {\textrm span}_{\Z} \{v_1,...,v_n\}$\footnote{Here we assume that the lattice $\Gamma$ has dimension $n$, i.e. $\{v_1,...,v_n\}$ is a set of linearly independent vectors in $\R^n$.} if and only if (after a translation)
\begin{align*}
	\Omega = \biggr\{ \sum_{j=1}^n t_j v_j : 0 < t_j < 1\quad \mbox{for any $1\leq j \leq n$.} \biggr\}.
\end{align*}\end{definition}

Let us also define the set of neighbouring reflections/translations of $\Omega$ as the set of reflections/translations for which the distance between the reflected/translated polytope and the original one is zero, i.e.,
\begin{align}
    &\mathcal{R}_\Omega^{nb} = \{ \sigma \in \mathcal{R}_\Omega : \overline{\sigma(\Omega)} \cap \overline{\Omega} \neq \emptyset \}, \quad \mathcal{T}_\Omega^{nb} = \{ v \in \Gamma : \overline{\Omega +v} \cap \overline{\Omega} \neq \emptyset \}. \label{eq:nbreflectiondef}
\end{align}
We can now construct the wave kernel in $\Omega$ explicitly. For this, it is helpful to introduce the reflection and translation of a function $g$, respectively, as
\begin{align*}
    \sigma^\#(g)(r) = g(\sigma r) \quad \mbox{for $\sigma \in \mathcal{R}_\Omega$}\quad\mbox{and}\quad \tau_v g(r) = g(r-v) \quad \mbox{for $v \in \Gamma$.}
\end{align*}
\begin{lemma}[Wave kernel on symmetric polytopes]\label{lem:wavekernelkaleidoscopic} Let $\Omega \subset \R^n$ be a strictly tessellating polytope or a fundamental domain of a lattice $\Gamma \subset \R^n$. Then, for any $g\in C^\infty_c(\Omega)$, the unique solution in $C^\infty(\overline{\Omega})$ to the initial value problem
\begin{align}
    \partial_{tt} u = \Delta_\Omega u, &\mbox{ in } \Omega \times (0,\infty) \quad \mbox{ with initial conditions } \begin{dcases}
    \partial_t u(r,0) = 0 \\
    u(r,0) = g(r) \\\end{dcases} \label{eq:initialboundaryproblem}
\end{align}
where $\Delta_\Omega$ is either the Dirichlet, Neumann, or periodic Laplacian, is given by
\begin{align}
    u(r,t) = \begin{dcases} \sum_{\sigma \in \mathcal{R}_\Omega} \det \sigma \bigr(E_0(t) \ast (\sigma^\# g)\bigr)(r) &\mbox{for Dirichet BCs,} \\\sum_{\sigma \in \mathcal{R}_\Omega}  \bigr(E_0(t) \ast (\sigma^\# g)\bigr)(r) &\mbox{for Neumann BCs,}\\
    \sum_{v \in \Gamma} \bigr(E_0(t) \ast (\tau_v g)\bigr)(r)  &\mbox{for periodic BCs,} \end{dcases} \label{eq:wavesolution}
\end{align}
where 
\begin{align*}
    E_0 \ast g(r) = \frac{1}{(2\pi)^n} \int_{\R^n} \cos (|k|t) \widehat{g}(k) e^{i k \scpr r} \,\mathrm{d}k
\end{align*}
is the wave kernel in $\R^n$ (see Appendix~\ref{sec:wavekernel}). 
\end{lemma}
\begin{proof} For simplicity, we prove only the Dirichlet case. (The other two are entirely analogous.) First, note that since $\mathrm{supp}(g) \subset \Omega$, by the strictly tessellating property \eqref{eq:tessellate},
\begin{align*}
    \mathrm{supp}(\sigma^\# g) \cap \mathrm{supp}(\tau^\# g) = \emptyset,
\end{align*}
for any $\sigma \neq \tau \in \mathcal{R}_\Omega$. In particular, $\sum_{\sigma \in \mathcal{R}_\Omega} \det \sigma \sigma^\# g$ is a sum of smooth functions with disjoint support and therefore smooth. Thus by Lemma \ref{lem:wavekernelRn}, the function $u(r,t)$ defined in \eqref{eq:wavesolution} is smooth and solves the wave equation in $\R^n$ with initial condition $u(r,0) = \sum_{\sigma \in \mathcal{R}_\Omega} \det \sigma \sigma^\# g$ and $\partial_t u(r,0) = 0$. Since uniqueness follows from the previous discussion, we just need to check that the boundary condition is satisfied. To this end, note that
\begin{align*}
    \sum_{\sigma \in \mathcal{R}_\Omega} \det \sigma \sigma^\# (\sigma_\ell^\# g) = \sum_{\sigma \in \mathcal{R}_\Omega} \det \sigma (\sigma_\ell \circ \sigma )^\# g = - \sum_{\sigma \in \mathcal{R}_\Omega} \det \sigma \sigma^\# g,
\end{align*}
where we used that $\sigma_\ell$ is invertible and $\det (\sigma \circ \sigma_\ell) = - \det \sigma$. Thus
\begin{align*}
    u(\sigma_\ell r,t) &= \sum_{\sigma \in \mathcal{R}_\Omega} \det \sigma  \sigma_\ell^\# \bigr(E_0  \ast (\sigma^\# g)\bigr)(r) =  \sum_{\sigma \in \mathcal{R}_\Omega} \det \sigma E_0(t) \ast\biggr((\sigma_\ell\circ \sigma)^\# g\biggr)(r) \\
    &= - u(r,t).
\end{align*}
To conclude, we note that $\sigma_\ell(r) = r$ for any $r\in F_\ell$ and $\partial \Omega = \bigcup_{\ell} F_\ell$, which implies that $u(r,t) = 0$ on $\partial \Omega$.\end{proof}
An useful corollary of the lemma above is the following generalized Poisson summation formula for radial functions.
\begin{corollary}[Generalized Poisson summation formula] \label{cor:generalizedpoisson} Let $\Omega\subset \R^n$ be a strictly tessellating polytope or a fundamental domain of a lattice $\Gamma$. Let $\Delta_\Omega$ be either the Dirichlet, Neumann, or periodic Laplacian in $\Omega$. Then, for any $f \in S(\R)$ even (i.e. $f(s) = f(-s)$ for any $s\in \R$), we have
\begin{align}
    \sum_{\lambda_j} f(\lambda_j) e_j(r)e_j(\tilde{r}) = \begin{dcases} \frac{1}{(2\pi)^n}\sum_{\sigma \in \mathcal{R}} \det \sigma \widehat{f(|\cdot|)}(r-\sigma \tilde{r}) &\mbox{for Dirichlet BCs,} \\
    \frac{1}{(2\pi)^n}\sum_{\sigma \in \mathcal{R}} \widehat{f(|\cdot|)}(r-\sigma \tilde{r}) &\mbox{for Neumann BCs,} \\
    \frac{1}{(2\pi)^n}\sum_{v \in \Gamma} \widehat{f(|\cdot|)}(r-\tilde{r}-v) &\mbox{for periodic BCs,} \\ \end{dcases} \label{eq:wavekernelid}
\end{align}
where $\widehat{f(|\cdot|)}$ is the Fourier transform of the radial function $r\in \R^n \mapsto f(|r|)$.
\end{corollary}
\begin{proof} First, observe that by the standard elliptic regularity estimate, for any $V\subset \subset U$, there exists some constant $C = C(V)>0$ such that
\begin{align*}
    \norm{e_j}_{W^{m,2}(V)} \leq C (1+\lambda_j)^{2m},
\end{align*}
for any $m \in \N$ and $\lambda_j$. Moreover, by the leading order Weyl law (see \eqref{weyl}), which can be shown to hold by the Dirichlet-Neumann bracketing technique \cite[Section XIII.15]{ReedSimonIV}, one can control the degeneracy of any eigenvalue by 
\begin{align*}
    d(\lambda_j) \coloneqq \dim \ker (-\Delta_\Omega-\lambda_j^2) \lesssim \lambda^n. 
\end{align*}
Thus from the classical Sobolev embedding we conclude that
\begin{align*}
    \sum_{\lambda_j \leq \lambda} |e_j(r)|^2 \lesssim (1+\lambda)^M ,
\end{align*}
for some $M \in \N$ and uniformly for $r\in V$. As a consequence, the left hand side of \eqref{eq:wavekernelid} is summable and the convergence is locally uniform in $\Omega\times \Omega$ as long as $f$ decays fast enough. Similarly, the right hand side of \eqref{eq:wavekernelid} is also an absolutely convergent sum, since $\widehat{f(|\cdot|)} \in S(\R^n)$ (as $f$ is even) and the set $\{\sigma r \}_{\sigma \in \mathcal{R}_\Omega}$ is uniformly discrete for any $r \in \Omega$. Finally, to obtain \eqref{eq:wavekernelid} we can integrate the right hand side of \eqref{eq:wavekernelid} against some test function $g\in C^\infty_c(\Omega)$ and use the identity \eqref{eq:prewavekernelid} with $u$ given by Theorem \ref{lem:wavekernelkaleidoscopic}. Then, we find
\begin{align*}
    \sum_{j}f(\lambda_j) e_j(r) &\inner{e_j, g}_{L^2(\Omega)} \\
    &= \frac{1}{2\pi}\int_{\R}\widehat{f}(t) \sum_{\sigma \in \mathcal{R}_\Omega} \det \sigma \frac{1}{(2\pi)^n}\int_{\R^n} \cos(t|k|) e^{ik\scpr r} \widehat{\sigma^\# g}(k) \,\mathrm{d}k \,\mathrm{d} t \\
    &= \sum_{\sigma \in \mathcal{R}_\Omega} \frac{\det \sigma}{(2\pi)^n} \int_{\R^n} f(|k|) e^{ik\scpr r} \widehat{\sigma^\# g}(k) \,\mathrm{d} k \\
    &=\sum_{\sigma \in \mathcal{R}_\Omega} \frac{\det \sigma}{(2\pi)^n} \int_{\R^n} \widehat{f(|\cdot|)}(r-\sigma \tilde{r}) g(\tilde{r}) \,\mathrm{d}\tilde{r}
\end{align*}
(where the change in the order of integration/summation can again be justified by the fast decay of $f$ and $g$). As the above identity holds for any test function $g\in C^\infty_c(\Omega)$, the result follows. \end{proof}

\section{Asymptotics of the spectral function} \label{sec:spectralfunction}

The goal of this section is to prove Theorems \ref{offdiagonalthm} and \ref{L2thm}. Throughout these proofs, we will often use some decaying properties of the Fourier transfoms of the $n$-dimensional ball and $(n-1)$-dimensional sphere. For later reference, we state these properties in the lemma below. (The reader can consult \cite{herz1962fourier} or \cite[Section 1.2]{sogge2017fourier} for a proof.)
\begin{lemma}[Fourier transform of the ball and sphere]\label{lem:fouriertransformsphere} Let $h_n(|k|) = \widehat{\chi_{B_1}}(k)/\omega_n$ be the normalized Fourier transform of the characteristic function of the unit ball in $\R^n$, and $\mu_n = \mathcal{H}^{n-1} \mres S^{n-1}$ be the $n-1$ Hausdorff measure restricted to the sphere $S^{n-1} = \{r\in \R^n : |r| = 1\}$. Then we have
\begin{align*}
    &|\partial^\alpha \widehat{\mu_n}(k)| \lesssim_{\alpha,n} \frac{1}{(1+|k|)^{\frac{n-1}{2}}} \quad \mbox{ and } \quad |\partial^\alpha \bigr(h_n(|k|)\bigr)| \lesssim_{\alpha,n} \frac{1}{(1+|k|)^{\frac{n+1}{2}}},
\end{align*}
where the implicit constant depends on $\alpha\in \N_0^n$ and $n\in \N$, but not on $k \in \R^n$.
\end{lemma}

\subsection{Uniform estimates}
We now present the proof of Theorem \ref{offdiagonalthm}. This proof is an adaptation of the arguments in \cite[Chapter 3]{sogge2014hangzhou}, where the diagonal version of Theorem \ref{offdiagonalthm} is proved for the periodic case.  

The first step in the proof is a uniform control on the growth of the sum of eigenfunctions (and its derivatives) in a small interval around $\lambda$.
\begin{lemma}[Sup-norm of Spectral $\epsilon$-Band]\label{lem:supnormcontrol} Let $\Omega$ be a strictly tessellating polytope or a fundamental domain of a lattice and $e_j$ be the eigenfunctions of the Laplacian under our usual BCs. Then, for any $\alpha \in \N_0^3$ and $1\leq \epsilon^{-1} \leq \lambda$, there exists a constant $C = C(\alpha)>0$ (independent of $\lambda$ and $\epsilon$) such that
\begin{align} \sum_{|\lambda_j-\lambda|\leq \epsilon} |\partial^\alpha e_j(r)|^2 \leq C \bigr(1+\lambda^{\frac{n-1}{2}+2|\alpha|}(\epsilon \lambda^{\frac{n-1}{2}} + \epsilon^{-\frac{n-1}{2}})\bigr) \quad \mbox{for any } r \in \Omega \label{degeneracycontrolv2} \end{align}
\end{lemma}
\begin{proof} The idea here is to estimate the sum in \eqref{degeneracycontrolv2} by studying the kernel of \[\eta_\lambda^{\epsilon}(\sqrt{-\Delta_\Omega})\] for some fast decaying non-negative function $\eta_{\lambda}^\epsilon$ that is positive in the interval $[\lambda-\epsilon,\lambda+\epsilon]$. To this end, let $\mu_n = \mathcal{H}^{n-1}\mres S^{n-1}$, and let $\eta \in S(\R)$ be a non-negative even function such that $\eta(s) >1$ for $|s| \leq 1$, and $\mathrm{supp}(\widehat{\eta})\subset [-1,1]$. Then, we define its even rescaled version by
\begin{align*}
    \eta^\epsilon_\lambda(\tau) \coloneqq \eta\biggr(\frac{\tau-\lambda}{\epsilon}\biggr) + \eta\biggr(\frac{\tau+\lambda}{\epsilon}\biggr),
\end{align*}
and note that $\mathrm{supp}(\widehat{\eta^\epsilon_\lambda})\subset [-\frac{1}{\epsilon},\frac{1}{\epsilon}]$. Thus from Lemma \ref{lem:fouriertransformsphere},
\begin{align}
    \partial^\alpha \widehat{\eta^\epsilon_\lambda(|\cdot|)}(z) &= \int_{0}^\infty \eta^\epsilon_\lambda(\tau) \partial^\alpha\bigr(\widehat{\mu_n}(\tau z)\bigr) \tau^{n-1} \,\mathrm{d} \tau \nonumber \\
    &= \epsilon \int_{0}^\infty \bigr(\eta(\tau-\lambda/\epsilon) +  \eta(\tau+\lambda/\epsilon)\bigr) \partial^\alpha\widehat{\mu_n}\bigr(\epsilon \tau z)(\epsilon \tau)^{n-1+|\alpha|} \,\mathrm{d} \tau \nonumber \\
    &\lesssim_\eta \epsilon \lambda^{\frac{n-1}{2}+|\alpha|}\min\biggr\{ \lambda^{\frac{n-1}{2}}, \frac{1}{|z|^{\frac{n-1}{2}}} \biggr\} . \label{eq:mainfouest}
\end{align}
Now, let us consider the set of reflections in $\mathcal{R}_\Omega$ for which the reflected polytope $\sigma(\Omega)$ lies at most a distance of $\frac{1}{\epsilon}$ away of the original polytope $\Omega$, i.e., 
\begin{align}
    \mathcal{R}_\epsilon\coloneqq \{ \sigma \in \mathcal{R}_\Omega : dist(\sigma(\Omega),\Omega) \leq \epsilon^{-1} \}. \label{eq:Reflectionepsilon}
\end{align}
Then, due to the strictly tessellating property, one can see that $\# \mathcal{R}_\epsilon \lesssim \frac{1}{\epsilon^n}$. Moreover, we claim that 
\begin{align}
    \bigr(\partial^\alpha \widehat{\eta^\epsilon_\lambda(|\cdot|)}\bigr)(r-\sigma \tilde{r}) = 0 \quad \mbox{ for any $\sigma \not \in \mathcal{R}_\epsilon, r,\tilde{r} \in \Omega$ and $\alpha\in \N_0^n$}. \label{eq:compactfourier}
\end{align} 
To show \eqref{eq:compactfourier}, just note that since $\mathrm{supp}(E_0(t)) \subset \{|r|\leq |t|\}$, we have
\begin{align*}
    \int_{\R^n} \widehat{\eta^\epsilon_\lambda(|\cdot|)}(\tilde{r}-r) g(r) \,\mathrm{d} r &= \int_{\R^n}\biggr(\int_{\R^n} \biggr(\frac{1}{\pi}\int_0^{\frac{1}{\epsilon}} \widehat{\eta^\epsilon_\lambda}(t) \cos(t|k|) \,\mathrm{d} t \biggr)e^{-ik\scpr (r-\tilde{r})} \,\mathrm{d} k\biggr) g(r) \,\mathrm{d} r\\
    &= \frac{(2\pi)^n}{\pi} \int_0^{\frac{1}{\epsilon}} \widehat{\eta^\epsilon_\lambda}(t) (E_0(t) \ast g)(\tilde{r}) \,\mathrm{d} t = 0,
\end{align*}
for any $g\in C^\infty_c(\R^n)$ with $dist(\mathrm{supp}(g),\tilde{r}) \geq \frac{1}{\epsilon}$. As $g$ was arbitrary, we conclude that $\widehat{\eta^\epsilon_\lambda}(r) = 0$ for any $|r|\geq \frac{1}{\epsilon}$ and \eqref{eq:compactfourier} holds. Hence, from Leibniz rule we have
\begin{align}
    \partial^\alpha_r \partial^\alpha_{\tilde{r}} \widehat{\eta^{\epsilon}_\lambda(|\cdot|)}(r-\sigma \tilde{r}) = \sum_{|\gamma| = |\alpha|} c_{\gamma,\sigma}\bigr(\partial^{\alpha+\gamma} \widehat{\eta^\epsilon_\lambda}(|\cdot|)\bigr)(r-\sigma \tilde{r}), \label{eq:leibniz}
\end{align}
where all $c_{\gamma,\sigma}$ are bounded by a constant independent of $\epsilon$, $\lambda$  and $\sigma$ (since all entries in the linear part of $\sigma$ are bounded by $1$). Therefore, by \eqref{eq:compactfourier}, Corollary \ref{cor:generalizedpoisson}, and estimate \eqref{eq:mainfouest} (and recalling that $\eta_\lambda^\epsilon \geq 1$ on $[\lambda-\epsilon,\lambda+\epsilon]$), we conclude that
\begin{align*}
    \sum_{|\lambda_j-\lambda|\leq \epsilon} |\partial^\alpha e_j(r)|^2 &\leq \sum_{\lambda_j} \eta^\epsilon_\lambda(\lambda_j) |\partial^\alpha e_j(r)|^2 \\
    &\leq \sum_{\sigma \in \mathcal{R}_\epsilon} \det \sigma \sum_{|\gamma| = |\alpha|} c_{\gamma,\sigma} \bigr|\bigr(\partial^{\alpha+\gamma}\widehat{\eta^\epsilon_\lambda(|\cdot|)}\bigr)(r-\sigma r) \bigr|\\
    &\lesssim \epsilon \lambda^{n-1+|\alpha|} + \epsilon \lambda^{\frac{n-1}{2}+|\alpha|}\sum_{1\leq dist(\sigma(\Omega),\Omega) \leq \frac{1}{\epsilon}}  \frac{1}{|r-\sigma r|^{\frac{n-1}{2}}} \\
    &\lesssim \lambda^{\frac{n-1}{2}+|\alpha|}\bigr(\epsilon \lambda^{\frac{n-1}{2}} +  \epsilon^{-\frac{n-1}{2}}\bigr).
\end{align*}\end{proof}

We can now complete the proof of Theorem \ref{offdiagonalthm}.

\begin{proof}[Proof of Theorem \ref{offdiagonalthm}] The idea here is similar to the previous proof; we choose a smooth version of the characteristic function of the interval $[-\lambda,\lambda]$ and use Lemma~\ref{lem:supnormcontrol} and the generalized Poisson summation to get the continuum version with error estimates controlled by powers of $\epsilon$ and $\lambda$. We can then estimate the error from smoothing the characteristic function and optimize $\epsilon$ to complete the proof.

Let $\chi_\lambda(s)$ be the characteristic function on the interval $[-\lambda,\lambda]$, and let $\eta \in S(\R)$ be an even nonnegative function with $\widehat{\eta}(0) = 1$ and $\mathrm{supp}(\widehat{\eta}) \subset [-1,1]$. In addition, let $\chi^\epsilon_\lambda$ be the mollification of $\chi_\lambda$ on the scale $\epsilon$, i.e., $\chi^\epsilon_\lambda(s) = \chi_\lambda \ast \bigr(\epsilon^{-1} \eta(\epsilon^{-1} \cdot)\bigr)(s)$, and $r^\epsilon_\lambda = \chi_\lambda - \chi^\epsilon_\lambda$ be the mollification error function. As $\eta$ decays fast, it is not hard to see that
\begin{align} |r_{\lambda}^\epsilon(s)|\lesssim_N \frac{1}{(1+\epsilon^{-1}|\lambda-s|)^{N}}+\frac{1}{(1+\epsilon^{-1}|\lambda+s|)^{N}} \lesssim_N \frac{1}{(1+\epsilon^{-1}|\lambda-s|)^{N}} \label{mollicontrol} \end{align}
for any $s\geq 0$. Thus denoting the mollified version of the spectral function by
\begin{align*}
    S^\epsilon_\lambda(r,\tilde{r}) = \sum_{\lambda_j} \chi^\epsilon_\lambda(\lambda_j) e_j(r) e_j(\tilde{r}),
\end{align*}
we can use \eqref{degeneracycontrolv2}, Cauchy-Schwarz and \eqref{mollicontrol} to bound the error with respect to $S_\lambda$ by
\begin{align} |\partial^\alpha_r\partial^\beta_{\tilde{r}} S_\lambda &- \partial^\alpha_r\partial^\beta_{\tilde{r}}S^\epsilon_\lambda| \\
&\leq \sum_{\ell=1}^\infty  \sum_{|\lambda_j-\ell \epsilon|\leq \epsilon} |r^\epsilon_\lambda(\lambda_j) \partial^\alpha e_j(r) \partial^\beta e_j(\tilde{r})| \nonumber \\
 &\lesssim_N \sum_{\ell} \frac{1}{(1+|\epsilon^{-1}\lambda - \ell|)^N} \biggr(\sum_{|\lambda_j-\ell\epsilon|\leq \epsilon} |\partial^\alpha e_j(r)|^2\biggr)^{\frac12}\biggr(\sum_{|\lambda_j-\ell\epsilon|\leq \epsilon} |\partial^\beta e_j(\tilde{r})|^2\biggr)^{\frac12}\nonumber \\
 &\lesssim \sum_{\ell}\frac{1+\ell^{n-1+|\alpha|+|\beta|} \epsilon^{n+|\alpha|+|\beta|} + \ell^{\frac{n-1}{2}+|\alpha|+|\beta|} \epsilon^{|\alpha|+|\beta|}}{(1+|\epsilon^{-1}\lambda - \ell||)^N} \nonumber \\
 &\lesssim \lambda^{\frac{n-1}{2}+|\alpha|+|\beta|}\bigr(\epsilon \lambda^{\frac{n-1}{2}} + \epsilon^{-\frac{n-1}{2}} \bigr)\quad \mbox{(for $\lambda$ big).}\label{mollyerror1} \end{align}
On the other hand, by applying Corollary \ref{cor:generalizedpoisson} to $S^\epsilon_\lambda$ and recalling from the last proof that $\widehat{\chi^\epsilon_\lambda(|\cdot|)}(k) = 0$ for $|k| \geq \frac{1}{\epsilon}$ (since $\mathrm{supp}(\widehat{\eta^\epsilon})\subset [-1/\epsilon,1/\epsilon]$), we find that
\begin{multline}
    \partial^\alpha_r \partial^\beta_{\tilde{r}} S^\epsilon_\lambda(r,\tilde{r}) =  \sum_{\sigma \in \mathcal{R}_\epsilon} \det \sigma \frac{1}{(2\pi)^n}\int_{\R^n} \bigr(\chi_\lambda(|k|) + r^\epsilon_\lambda(|k|)\bigr) \partial^\alpha_r \partial^\beta_{\tilde{r}}\bigr(e^{ik \scpr(r-\sigma \tilde{r})}\bigr)\,\mathrm{d}k \nonumber \\
    = \sum_{\sigma \in \mathcal{R}_\epsilon} \frac{\det \sigma}{(2\pi)^n}\biggr(\omega_n\lambda^n \partial^\alpha_r \partial^\beta_{\tilde{r}} h_n(\lambda |r-\sigma \tilde{r}|) + \int_0^\infty r^\epsilon_\lambda(\tau) \tau^{n-1} \partial^\alpha_r \partial^\beta_{\tilde{r}}\widehat{\mu_n}\bigr(\tau(r-\sigma \tilde{r})\bigr)) \,\mathrm{d} \tau\biggr).\label{eq:ctmplusrest}
\end{multline}
Moreover, from \eqref{mollicontrol} and Lemma \ref{lem:fouriertransformsphere} we have
\begin{align*}
    r^\epsilon_\lambda(\tau) \tau^{n-1} \partial^\alpha_r \partial^\beta_{\tilde{r}}\widehat{\mu_n}\bigr(\tau(r-\sigma \tilde{r})\bigr)\lesssim \frac{\tau^{\frac{n-1}{2}+|\alpha|+|\beta|}}{(1+\epsilon^{-1}|\lambda-\tau|)^N} \min\{ \tau^{\frac{n-1}{2}}, |r-\sigma \tilde{r}|^{-\frac{n-1}{2}}\}.
\end{align*}
By integrating the estimate above over $(0,\infty)$ and summing over $\sigma \in \mathcal{R}_\epsilon$, we can see that the last term in \eqref{eq:ctmplusrest} yields (at most) an error of order $\mathcal{O}\bigr(\lambda^{\frac{n-1}{2}+|\alpha|+|\beta|}(\epsilon \lambda^{\frac{n-1}{2}} + \epsilon^{-\frac{n-1}{2}})\bigr)$. Therefore, we conclude from \eqref{mollyerror1}, \eqref{eq:ctmplusrest}, and the decay of $h_n$ that
\begin{align}
    \partial^\alpha_r \partial^\beta_{\tilde{r}} S_\lambda(r,\tilde{r}) &= \partial^\alpha_r \partial^\beta_{\tilde{r}}S_\lambda^{\textnormal{ctm}}(r,\tilde{r}) + \sum_{\sigma \in \mathcal{R}_\epsilon\setminus \mathcal{R}_1} \det \sigma \frac{\omega_n}{(2\pi)^n}  \underbrace{ \lambda^n \partial^\alpha_r \partial^\beta_{\tilde{r}} h_n(\lambda |r-\lambda \tilde{r}|)}_{\lesssim \lambda^{\frac{n-1}{2}+|\alpha|+|\beta|}|r-\sigma \tilde{r}|^{-\frac{n+1}{2}}} \nonumber \\
    &+\mathcal{O}\bigr(\lambda^{\frac{n-1}{2}+|\alpha|+|\beta|}(\epsilon \lambda^{\frac{n-1}{2}} + \epsilon^{-\frac{n-1}{2}})\bigr) \nonumber \\
    &=\partial^\alpha_r \partial^\beta_{\tilde{r}}S_\lambda^{\textnormal{ctm}}(r,\tilde{r}) + \mathcal{O}\bigr(\lambda^{\frac{n-1}{2}+|\alpha|+|\beta|}(\epsilon \lambda^{\frac{n-1}{2}} + \epsilon^{-\frac{n-1}{2}})\bigr).
\end{align}
The result now follows by setting $\epsilon = \lambda^{-\frac{n-1}{n+1}}$. The proof for the periodic and Neumann cases is a straightforward adaptation of the arguments presented above. \end{proof}

\subsection{$L^2$ estimate}

We now turn to the $L^2$ estimates for the spectral function. This result can be seen as a quantified version of the $L^2$ convergence of the  Wigner transform of the normalized spectral function in the work by Schmidt \cite[Theorem 1.2]{schmidt2011localized}. However, unlike the more classical (and more general) methods in \cite{schmidt2011localized}, our proof is again based on the wave kernel constructed before.

\begin{proof}[Proof of Theorem \ref{L2thm}] As in the proof of Theorem~\ref{offdiagonalthm}, we let $\chi_\lambda$ be the characteristic function on the interval $[-\lambda,\lambda]$ and $\eta \in S(\R)$ be an nonnegative even function with $\widehat{\eta} =1$ on a neighbourhood of $0$. Then, we define the mollified version of $\chi_\lambda$, the mollifying error function, and the smoothed spectral function as $\chi^1_\lambda \coloneqq \chi_\lambda \ast \eta$, $r_\lambda \coloneqq \chi^1_\lambda- \chi_\lambda$, and $S^1_\lambda = \sum_{j} \chi^1_\lambda(\lambda_j)e_j(r)\overline{e_j}(\tilde{r})$, respectively. Hence, by the orthogonality of $e_j$, we have
\begin{align*} 
    \norm{S_\lambda - S^1_\lambda}_{L^2(\Omega\times\Omega)}^2 &= \sum_{j,k} r_\lambda(\lambda_j) \overline{r}_\lambda(\lambda_k)\int_{\Omega\times \Omega} (e_j\overline{e_k})(r) (\overline{e_j}e_k)(\tilde{r}) \,\mathrm{d}r \,\mathrm{d}\tilde{r} \\
    &\lesssim \sum_{j} |r_\lambda(j)|^2 (N(j+1)-N(j)) \\
    &\lesssim \sum_{j=1} (1+(\lambda-j))^{-N} j^{n-1} \lesssim \lambda^{n-1}.
\end{align*}
So up to an error $\lesssim \lambda^{\frac{n-1}{2}}$, we can work with the smoothed spectral function $S_\lambda^1$. Now, since we do not vary the support of $\widehat{\eta}$ in this proof (no scaling with $\epsilon$), we see that $\widehat{\chi^1_\lambda} = \widehat{\chi}_\lambda \widehat{\eta}$ has support on a fixed neighbourhood of $0$. In particular, if we choose the support of $\widehat{\eta}$ small enough and apply the generalized Poisson summation in Corollary \ref{cor:generalizedpoisson} to $\chi^1_\lambda$, we conclude (see \eqref{eq:compactfourier} in the previous proof) that all terms with $\sigma \in \mathcal{R}_\Omega \setminus \mathcal{R}_\Omega^{nb}$ vanish. Therefore, the result follows if we show that for any $\sigma \in \mathcal{R}_\Omega$ the following estimate holds:
\begin{align*}
    \norm{\widehat{\chi_\lambda^1(|\cdot|)}(r-\sigma \tilde{r}) - \widehat{\chi_\lambda(|\cdot|)}(r-\sigma \tilde{r})}_{L^2(\Omega \times \Omega)}  = \norm{\widehat{r_\lambda(|\cdot|)}(r-\sigma \tilde{r})}_{L^2(\Omega\times \Omega)} \lesssim \lambda^{\frac{n-1}{2}},
\end{align*}
where $\widehat{g(|\cdot|)}$ is the Fourier transform in $\R^n$ of the function $r \mapsto g(|r|)$. This estimate is a direct consequence of Plancherel's theorem and the estimate $r_\lambda(|r|) \leq (1+|\lambda- |r||)^{-N}$. \end{proof}

\begin{remark*} Note that we only used the wave kernel for times of order $1$ here\footnote{Unlike in the $L^\infty$ case, we could not use the large times wave kernel to improve the remainders in the $L^2$ case.} . In particular, the same estimate is expected to hold on more general domains (e.g. smooth ones). \end{remark*}

We can now interpolate between the $L^2$ and $L^\infty$ estimate to obtain
\begin{corollary}[$L^p$ estimates]\label{Lpthm} Let $\Omega$ be a strictly tessellating polytope or a fundamental domain of a lattice, and let $S_\lambda$ be the spectral function of the periodic, Dirichlet or Neumann Laplacian in $\Omega$. Then,
\begin{align} \norm{S_\lambda-S^{\textnormal{ctm}}_\lambda}_{L^p(\Omega\times \Omega)} \lesssim \lambda^{(n-1)\left(1-\frac{1}{p}\right) - \frac{n-1}{n+1} \left(1-\frac{2}{p}\right)}\label{Lpmain}, \end{align}
where $S^{\textnormal{ctm}}_\lambda$ is the continuum spectral function defined in Theorem \ref{offdiagonalthm}.
\end{corollary}

\section{Asymptotics of functionals} \label{sec:asympexchangefunctionals} \label{sec:exchangepoly}
In this section we present the proof of the main results. For these proofs, we shall use two geometric lemmas.

The first lemma is a lower bound on the distance between points in the original polytope and points in the reflected one. To state this lemma, let us introduce some more notation. First recall that, since $\Omega$ is an open convex polytope with faces $\{F_j \}_{j\leq m}$, there exists $\{\alpha_j\}_{j\leq m} \subset \R$ such that 
\begin{align}
	\Omega = \{ r \in \R^n : r \scpr n_j > \alpha_j \mbox{ for any $1\leq j \leq m$} \}, \label{eq:polytope}
\end{align}
where $n_j$ is the unit inward-pointing normal vector to the face $F_j$. Moreover, for any $\sigma \in \mathcal{R}_\Omega^{nb}$ there exists $\{j_1,...,j_p\} \subset \{1,...,m\}$ such that $\overline{\Omega}\cap \overline{\sigma(\Omega)} = \bigcap_{k=1}^p F_{j_k}$ and the interior
\begin{align}
{\textrm int} \bigcap_{k=1}^p F_{j_k} \coloneqq \biggr\{ r\in \R^n: r\scpr n_j \begin{dcases} = \alpha_j \quad &\mbox{if $j \in \{j_k\}_{k\leq p}$,}\\
	> \alpha_j &\mbox{otherwise} \end{dcases} \biggr\}  \label{eq:interior}
\end{align}
is non-empty (see Lemma~\ref{lem:faces} below). We then denote the metric projection along the affine space extending this intersection by $\pi_\sigma$, i.e., 
\begin{align}
	\pi_\sigma r = {\textrm argmin} \{ |r-r'| :  r'\in \R^n \quad\mbox{and}\quad n_{j_k} \scpr r' = \alpha_{j_k} \quad\mbox{for all $1\leq k \leq p$}\}. \label{eq:projsigmadef}
\end{align}  
We also define the complementary projection as $\pi_\sigma^\perp r \coloneqq r- \pi_\sigma r$. 
\begin{lemma}[Lower bound on reflected distances]\label{lem:reflectionbound} Let $\sigma \in \mathcal{R}^{nb}_{\Omega}$, then
\begin{align*}
     |r-\sigma r'| \gtrsim  |\pi_\sigma r - \pi_\sigma r'| + |\pi_\sigma^\perp r + \pi_\sigma^\perp r'|\quad \mbox{ and } \quad |r-\sigma r'|\gtrsim |r-r'|
\end{align*}
for any $r, r' \in \Omega$. (With the convention that $\pi_\sigma(r) = r$ if $\sigma$ is the identity.)
\end{lemma}
\begin{proof} After relabelling the faces and translating our reference frame, we can assume that $0 \in  \bigcap_{j=1}^p F_j = \overline{\Omega} \cap \overline{\sigma(\Omega)}$. In this case, $\sigma$ is a linear transformation given by some composition of the (linear) reflections $\{\sigma_j\}_{j\leq p}$ (see Lemma~\ref{lem:faces} below) and $\pi_\sigma$ becomes the orthogonal projections along the subspace 
\begin{align}
V_\sigma = \{ r \in \R^n : r\scpr n_j = 0\quad \mbox{for all $j \leq p$} \}.
\end{align}
In particular, $\sigma r = r$ for any $r\in V_\sigma$ and $\sigma r \in V_\sigma^\perp$ for any $r\in V_\sigma^\perp$. If we now define the closed conic sets $C_\Omega = \{ r \in  V_\sigma^\perp : r \scpr n_j \geq 0 \mbox{ for } 1\leq j \leq p\}$ and $\sigma(C_\Omega) = \{ \sigma r \in V_\sigma^\perp : r \in C_\Omega\}$, then we have
\begin{align}
    \overline{\Omega} \subset V_\sigma \oplus C_\Omega \quad \mbox{ and } \quad \overline{\sigma(\Omega)} \subset V_\sigma \oplus \sigma(C_{\Omega}). \label{eq:conicinclusions}
\end{align}
Moreover, one can show that $C_\Omega \cap \sigma(C_{\Omega}) = \{0\}$. Indeed, if $r,\widetilde{r} \in C_\Omega$ with $r = \sigma \tilde{r}$, then for any $p \in {\textrm int} \bigcup_{k=1}^p F_k$ (see \eqref{eq:interior}) we have $\delta r + p = \sigma(\delta \tilde{r} + p) \in \overline{\Omega} \cap \overline{\sigma(\Omega)} \subset V_\sigma$ for $\delta>0$ small enough, which implies that $r = \tilde{r} = 0$. Thus $C_\Omega$ and $\sigma(C_\Omega)$ are closed conic subsets that intersect only at zero. Consequently,
\begin{align*}
    |r-\sigma r'| \gtrsim |r| + |\sigma r'| \quad \mbox{for any $r,r' \in C_\Omega,$} 
\end{align*}
where the implicit constant is independent of $r$ and $r'$. From this inequality, the inclusions in \eqref{eq:conicinclusions}, and the fact that $V_\sigma$ is invariant under $\sigma$, we conclude that
\begin{align}
    |r-\sigma r'|^2 & = |\pi_\sigma(r-r')|^2 + |\pi_\sigma^\perp (r-\sigma r')|^2  = |\pi_\sigma (r-r')|^2 + |\pi_\sigma^\perp r-\sigma  \pi_\sigma^\perp r'|^2 \nonumber \\
    &\gtrsim |\pi_\sigma (r-r')|^2 + (|\pi_\sigma^\perp r|+|\pi_\sigma^\perp \sigma r'|)^2\quad\mbox{for any $r, r' \in \Omega$.}\label{eq:finalreflectest}
\end{align}
Lemma \ref{lem:reflectionbound} now follows from \eqref{eq:finalreflectest}, the triangle inequality, and the fact that $\sigma$ is an isometry. \end{proof}
The second geometric lemma we need is a first-order Taylor expansion of the function $w \mapsto |(\Omega-w) \cap \Omega|$ at $w=0$.
\begin{lemma}[Distributional derivative of $\Omega \cap (\Omega-w)$]\label{lem:overlap} Let $\Omega \subset \R^n$ be a polytope. Then, for any $a \in C^\infty_c(\R^n)$, there exists a constant $C = C(|\Omega|,|\partial \Omega|, \norm{a}_{L^\infty}, \norm{\nabla a}_{L^\infty}) >0$ such that for any $w \in \R^n$
\begin{align} \biggr|\int_{\Omega \cap (\Omega -w)} a(r) \,\mathrm{d}r - \int_{\Omega} a(r) \,\mathrm{d} r + \int_{\partial \Omega} a(r) (n(r) \scpr w)_+ \,\mathrm{d}\mathcal{H}^{n-1}(r) \biggr|\leq C |w|^2, \label{overlapping} \end{align}
where $n(r)$ is the outward-pointing unit normal and $f(r)_+ \coloneqq \max\{f(r),0\}$. In particular, we have
\begin{multline}
    \int_{(\Omega-w_1)\cap (\Omega+w_2)} a(r) \,\mathrm{d} r \\
    = \int_\Omega a(r) \,\mathrm{d}r - \int_{\partial \Omega}a(r) \frac{|n(r)\scpr (w_1+w_2)| + n(r) \scpr(w_1-w_2)}{2} \,\mathrm{d} \mathcal{H}^{n-1}(r)   \\
    + \mathcal{O}(|w_1|^2+|w_2|^2) . \label{eq:overlapest1}
\end{multline}\end{lemma}

\begin{proof} Since $\Omega$ is bounded, it is clear that $F(z) = \int_{\Omega \cap (\Omega-z)} a(r) \,\mathrm{d}r$ is continuous and compactly supported. Therefore, it is enough to show that \eqref{overlapping} holds on a neighbourhood of $0$. For this, let us define the sets
\begin{align*}
    F_k(z) \coloneqq \{ r \in \Omega: \alpha_k - (n_k \scpr z)_+\leq n_k \scpr r \leq \alpha_k \}, 
\end{align*}
where $\alpha_k$ and $n_k$ are the same from \eqref{eq:polytope}. Then we find that $\Omega\setminus (\Omega-z) = \bigcup_{k=1}^m F_k(z)$ and $|F_k(z) \cap F_j(z)| = \mathcal{O}(|z|^2)$ for $j \neq k$. Thus,
\begin{align}
    \int_{(\Omega-z)\cap \Omega} a(r) \,\mathrm{d}r - \int_{\Omega} a(r) \,\mathrm{d}r +\sum_{k=1}^m\int_{F_k(z)} a(r) \,\mathrm{d} r + \mathcal{O}(\norm{a}_{L^\infty}|z|^2) \label{eq:bulkest}
\end{align}
Next, note that since $\Omega$ is a convex polytope, up to an error $\lesssim \norm{a}_{L^\infty}|z|^2$, we can replace the integration over the set $F_k(z)$ by integration over the set $\{ r- \tau n_j : r \in F_k ,  0\leq \tau \leq (n(r)\scpr z)_+\}  \cong F_k \times[0,(n_k\scpr z)_+]$. Therefore, we find that
\begin{align*}
    \int_{F_k(z)} a(r) \,\mathrm{d}r &= \int_{F_k} \,\mathrm{d}\mathcal{H}^{n-1}(r) \int_0^{(n(r)\scpr z)_+} a\bigr(r - \tau n(r)\bigr) \,\mathrm{d} \tau  + \mathcal{O}(\norm{a}_{L^\infty}|z|^2) \\
    &= \int_{F_k} a(r) (n(r)\scpr z)_+ \,\mathrm{d}\mathcal{H}^{n-1}(r) + \mathcal{O}\bigr(\norm{|a|}_{L^\infty}|z|^2 + \norm{\nabla a}_{L^\infty}|z|^2\bigr),
\end{align*}
which together with \eqref{eq:bulkest} completes the proof.
\end{proof}

\begin{remark*} Note that Lemma \ref{lem:overlap} also holds for smooth domains by taking a partition of the unity along the boundary. \end{remark*}

\subsection{Proof of Theorem \ref{semilocalthm}}

Throughout this section, we use $\nu_\lambda$ for the combined function
\begin{align}
    \nu_\lambda(r) = 2(S_{s,\lambda}(r), \nabla S_{s,\lambda}(r) ) \in \R^{1+n} ,
\end{align}
where $S_{s,\lambda}$ and $\nabla S_{s,\lambda}$ are the re-scaled spectral function and its gradient. Similarly, the continuum version $\nu^{\textnormal{ctm}}_\lambda(r)$ is defined by using the continuum spectral function
\begin{align}
    S_{s,\lambda}^{\textnormal{ctm}}(r) = \frac{2\omega_n}{(2\pi)^n} \biggr( 1 - \sum_{\substack{\sigma \in \mathcal{R}_{\Omega_\lambda}^{nb} \setminus \{id\}}} \det \sigma h_n(|r-\sigma r|)\biggr).
\end{align}
We start with the asymptotic expansion of 
\begin{align}
    F^{\textnormal{ctm}}(\lambda) = \int_{\Omega_\lambda} f\bigr(\nu^{\textnormal{ctm}}_\lambda(r)\bigr)\,\mathrm{d}r . \label{eq:ctmsemilocal}
\end{align}
\begin{lemma}[Continuum semi-local asymptotics]\label{weyl2} Let \[f \in C^1((0,\infty)\times \R^n) \cap L^\infty_{\mathrm{loc}}([0,\infty)\times \R^n),\] 
then we have
\begin{align*}
    F^{\textnormal{ctm}}(\lambda) = \int_{\Omega_\lambda} f(\nu^{\textnormal{ctm}}_\lambda(r)) \,\mathrm{d} r = f(\nu_0) |\Omega| \lambda^n + c(f,\Omega) \lambda^{n-1} + \smallO(\lambda^{n-1}),
\end{align*}
where $\nu_0$ and the coefficient $c(f,\Omega)$ are defined in Theorem \ref{semilocalthm}. 
\end{lemma}
\begin{proof} First, we want to use the $C^1$ regularity of $f$ to estimate the difference $F^{\textnormal{ctm}}(\lambda) - f(\nu_0)|\Omega_\lambda|$. Since $f(a,b)$ is only $C^1$ at the points $a>0$, we start by showing that $S^{\textnormal{ctm}}_{s,\lambda}$ only vanishes close to the edges and faces of $\Omega$. For this, first note that $h_n(\tau) =1$ if and only if $\tau =0$ and that $h_n(\tau) \ra 0$ as $\tau \ra \infty$. Therefore, for any $\delta >0$ we can find $C_0, C(\delta), c(\delta) >0$ such that
\begin{align} S^{\textnormal{ctm}}_{s,\lambda}(r) > c(\delta) \quad \mbox{ and }\quad |\nabla \nu^{\textnormal{ctm}}_\lambda(r)| \leq C_0 , \label{eq:ctmdensitybound}
\end{align}
for any $r$ in the set
\begin{align}
    \Omega^{\delta}_\lambda \coloneqq \{ r \in \Omega_\lambda : \min_{1\leq \ell \leq m}\{| r- \sigma_\ell (r)|\} \geq \delta \mbox{ and } \min_{\sigma \in \mathcal{R}_{\Omega_\lambda}^{nb}\setminus\{\sigma_\ell\}_{0\leq \ell \leq m}}\{|r-\sigma r|\} \geq C(\delta) \}, \label{eq:Omegadeltadef}
\end{align}
where $\sigma_{\ell}$ is the reflection over the re-scaled face $\lambda F_\ell$ of the re-scaled polytope $\Omega_\lambda$ and $\sigma_0$ is the identity on $\R^n$. In other words, $\Omega_\lambda^\delta$ is the set of points of $\Omega_\lambda$ which are at least a distance $\delta$ of the faces and a distance of order $C(\delta)$ of the edges of $\Omega_\lambda$ (see Lemmas~\ref{lem:reflectionbound} and \ref{lem:faces}). So from \eqref{eq:ctmdensitybound}, the assumptions on $f$, and the simple estimate
\begin{align*}
    |\Omega_\lambda \setminus \Omega_\lambda^{\delta}| \lesssim C(\delta)^2 \lambda^{n-2} + \delta \lambda^{n-1}, 
\end{align*}
we find that 
\begin{align}
    F(\lambda) - f(\nu_0)|\Omega_\lambda| &=  \int_{\Omega^\delta_\lambda} \int_0^1 \nabla f\bigr(\nu_0 + t(\nu_\lambda^{\textnormal{ctm}}(r)-\nu_0)\bigr) \scpr (\nu^{\textnormal{ctm}}_\lambda(r)-\nu_0) \,\mathrm{d} t \,\mathrm{d} r  \nonumber \\
    & +\mathcal{O}(C(\delta)^2 \lambda^{n-2} + \delta \lambda^{n-1}). \label{eq:calculusest}
\end{align}
The next step is to expand the difference $\nu_\lambda^{\textrm ctm}-\nu_0$ that appears outside $\nabla f$ in a sum of terms over $\mathcal{R}_\Omega^{nb} \setminus \{\sigma_0\}$, and then get rid of the terms that only give lower order contributions. To this end, let us define
\begin{align*}
    \rho_\sigma(r) \coloneqq \frac{2\omega_n}{(2\pi)^n} h_n(|r-\sigma r|) \quad \mbox{ and } \quad \nu_\sigma \coloneqq (\rho_\sigma, \nabla \rho_\sigma).
\end{align*} 
Then since ${\textrm range}(\pi_\sigma)$ is an affine subspace of dimension at most $n-2$ for any $\sigma \in \mathcal{R}_{\Omega_\lambda}\setminus \{\sigma_\ell\}_{1\leq \ell \leq m}$, we can use Lemma \ref{lem:reflectionbound}, the decay of $h_n$, and the local boundedness of the gradient of $f$ to show that
\begin{align}
    \int_{\Omega^\delta_\lambda} \int_0^1 \nabla f\bigr(\nu_0 + t(\nu^{\textnormal{ctm}}_\lambda-\nu_0\bigr) \scpr \nu_\sigma(r) \,\mathrm{d} r \lesssim_\delta \int_{\Omega^\delta_\lambda}(1+|\pi_\sigma^\perp r|)^{-\frac{n+1}{2}}\,\mathrm{d} r \lesssim_\delta \lambda^{n-\min\{\frac{n+1}{2},2\}} \label{eq:neglectest}
\end{align}
for any $\sigma \in \mathcal{R}^{nb}_{\Omega_\lambda} \setminus \{\sigma_{\ell}\}_{0\leq \ell \leq m}$. As a consequence, we are left with the terms
\begin{align*}
    K_{\ell}(\lambda,\delta) \coloneqq \int_{\Omega^\delta_\lambda} \int_0^1 \nabla f\bigr(\nu_0 + t(\nu^{\textnormal{ctm}}_\lambda(r) - \nu_0)\bigr)\scpr \bigr(-\nu_{\sigma_\ell}(r)\bigr) \,\mathrm{d} t\,\mathrm{d} r  \quad \mbox{ for $1 \leq \ell \leq m$.}
\end{align*}
To obtain the asymptotics of $K_\ell$, we can assume (without loss of generality) that the face $F_{\ell}$ lies on the plane $\{r \in \R^n : r_n = 0\}$ and the inward-pointing normal is $n_{\ell} = (0,....,1)$. Under this assumption, $\rho_{\ell}(r) = 2\omega_n/(2\pi)^n h_n(2r_n)$ and 
\begin{align*}
    \nu_{\sigma_\ell}(r) =  \frac{2\omega_n}{(2\pi)^n} \bigr(h_n(2r_n), 2n_\ell \dot{h_n}(2r_n)\bigr).
\end{align*}
Moreover, one can check that
\begin{align}
    &\lim_{\lambda \ra \infty} \chi_{\Omega^\delta_\lambda}(\lambda r_1,...,\lambda r_{n-1}, r_n) = \chi_{F_\ell}(r_1,...,r_{n-1},0)\chi_{(\delta,\infty)}(r_n) \quad\mbox{and} \\
    &\lim_{\lambda \ra \infty} \nu_\lambda^{\textrm ctm}(\lambda r_1,...,\lambda r_{n-1},r_n)-\nu_0 = - \nu_{\sigma_\ell}(r_n) 
\end{align}
for almost every $r \in \R^{n-1} \times (0,\infty)$, where $\chi_A$ stands for the characteristic function of the set $A$. Thus since $\nu_{\sigma_\ell}(r) \lesssim (1+|r_n|)^{-\frac{n+1}{2}} \in L^1(\R)$, we can now re-scale the variables $r_1,...,r_{n-1}$ by $\lambda$ and apply the dominated convergence theorem to conclude that
\begin{align}
    \lim_{\lambda \ra \infty}&\frac{K_\ell(\lambda,\delta)}{\lambda^{n-1}} \notag \\
    &=\int_\delta^\infty \int_{\R^{n-1}} \chi_{F_\ell}(r_1,..,r_{n-1}) \int_0^1 \nabla f\bigr(\nu_0 - t \nu_{\sigma_\ell}(r_n))\scpr \bigr(- \nu_{\sigma_\ell}(r_n)\bigr) \,\mathrm{d} t  \,\mathrm{d}r_1...\,\mathrm{d} r_n \nonumber \\
    & = \int_{F_\ell}  \int_0^\infty f\bigr(\nu_0-t\nu_1(r_n,r')\bigr) - f(\nu_0) \,\mathrm{d} r_n \,\mathrm{d} \mathcal{H}^{n-1}(r') +\mathcal{O}(\delta), \label{eq:limitest}
\end{align}
where $\nu_1(r_n,r') = \frac{2\omega_n}{(2\pi)^n} \bigr(h(2r_n), 2 n(r') \dot{h}(2r_n)\bigr)$. The proof now follows by plugging \eqref{eq:neglectest} and  \eqref{eq:limitest} in \eqref{eq:calculusest} and taking the limit $\lambda \ra 0$ and then $\delta \ra 0$. \end{proof}

To complete the proof of Theorem \ref{semilocalthm}, it is enough to show that
\begin{align}
    F^{\textnormal{ctm}}(\lambda) - F(\lambda)  = \smallO(\lambda^{n-1}). \label{eq:endest}
\end{align}
So fix again some $\delta>0$ and let $\Omega_\lambda^\delta$ be defined as in \eqref{eq:Omegadeltadef}. Then from \eqref{eq:ctmdensitybound} and Theorem \ref{offdiagonalthm} we find that
\begin{align}
    S_{s,\lambda}(r) \geq c(\delta)/2 \mbox{ in } \Omega_\lambda^\delta \quad \mbox{ and } \quad |\nabla \nu_\lambda(r)| \leq 2C_0 \mbox{ in } \Omega_\lambda \label{eq:unifbounded}
\end{align}
for $\lambda$ big enough. It thus follows from the assumptions on $f$ and Theorem \ref{offdiagonalthm} that
\begin{align*}
    \int_{\Omega_\lambda} f(\nu_\lambda) - f(\nu^{\textnormal{ctm}}_\lambda) \,\mathrm{d} r &\lesssim |\Omega_\lambda \setminus \Omega_\lambda^{\delta}| + D(\delta) \int_{\Omega_\lambda^{\delta}} |\nu^{\textnormal{ctm}}(r)_\lambda - \nu_\lambda(r)| \,\mathrm{d} r \\
    &\lesssim C(\delta)^2 \lambda^{n-2} + \delta \lambda^{n-1} + D(\delta) \lambda^{n-1-\frac{n-1}{n+1}} \quad \mbox{for some $D(\delta)>0$.}
\end{align*}
Therefore, we can divide the estimate above by $\lambda^{n-1}$, send $\lambda \ra \infty$ and then $\delta \ra 0$ to obtain \eqref{eq:endest}.

\subsection{Proof of Theorem \ref{exchangethm}}

As in the previous section, we only work out the Dirichlet case in detail. We comment on the modifications necessary for the Neumann and periodic cases at the end of the proof. We start again by computing the asymptotics of the exchange energy for the continuum spectral function
\begin{align}
    E_x^{\textnormal{ctm}}(\lambda) &= \int_{\Omega^2} \frac{|S^{\textnormal{ctm}}_\lambda|^2}{|r-r'|} \,\mathrm{d}r\,\mathrm{d}r'\notag \\
    &= \frac{\omega_n^2 \lambda^s}{(2\pi)^{2n}} \sum_{\sigma, \tau \in \mathcal{R}^{nb}_{\Omega_\lambda}}\det \sigma \tau  \underbrace{\int_{\Omega_\lambda^2}\frac{h_n(|r-\sigma r'|)h_n(|r-\tau r'|)}{|r-r'|^s} \,\mathrm{d}r \,\mathrm{d}r'}_{\coloneqq E_{\sigma,\tau}(\lambda)}.  \label{eq:exchangeenergysplitting}
\end{align}
The first step here is to get rid of the terms $E_{\sigma,\tau}(\lambda)$ that only gives lower order contributions; to this end, we use the following Lemma.
\begin{lemma}[Lower order contribution]\label{lem:lowerorder} Suppose that either $\sigma \not \in \{\sigma_{\ell}\}_{0\leq \ell \leq m}$ or $(\sigma,\tau) = (\sigma_j,\sigma_k)$ where $1\leq j \neq k \leq m$. Then we have
\begin{align}
    \lambda^s E_{\sigma,\tau}(\lambda) = \mathcal{O}(\lambda^{\max\{n-2+s,n-1,\frac{n-1}{2}+s\}+\epsilon}), \label{eq:edgerest}
\end{align}
for any $\epsilon>0$.
\end{lemma}
\begin{proof} The key idea is to split the decay of $h_n$ over linear combinations of the components of $r$ and $r'$ in order to compensate for the integration in $\Omega_\lambda \times \Omega_\lambda$ in as many directions as possible. So first, from Lemma \ref{lem:reflectionbound} we have
\begin{align*}
    E_{\sigma,\tau}(\lambda) \lesssim \int_{\Omega_\lambda^2} \frac{(1+|r-r'|)^{-\frac{n+1}{2}}}{(1+|\pi_\sigma r - \pi_\sigma r'|)^{\frac{n+1}{2}-x}(1+|\pi_\sigma^\perp r + \pi_\sigma^\perp r'|)^{x}} \frac{1}{|r-r'|^s} \,\mathrm{d}r\,\mathrm{d}r',
\end{align*}
for any $0\leq x\leq (n+1)/2$. Hence, identifying the spaces ${\textrm range}(\pi_\sigma)\approx \R^d$ and $\textnormal{range}(\pi_\sigma^\perp) \approx \R^{n-d}$, we can make the change of variables \[(z,z',w,w') = (\pi_\sigma r - \pi_\sigma r', \pi_\sigma^\perp r - \pi_\sigma^\perp r', \pi_\sigma^\perp r + \pi_\sigma^\perp r', \pi_\sigma r + \pi_\sigma r') \in \R^d \times \R^{n-d}\times \R^{n-d} \times \R^d\] to find that
\begin{align*}
    \lambda^s E_{\sigma,\tau}(\lambda) &\lesssim \lambda^s \int_{\substack{|z|+|z'|\lesssim \lambda \\ |w|+|w'|\lesssim \lambda}}(1+|z|)^{-\frac{n+1}{2}+x}(1+|w|)^{-x}(|z|+|z'|)^{-\frac{n+1}{2}-s}\,\mathrm{d}z'\,\mathrm{d}z\,\mathrm{d}w'\,\mathrm{d}w \\
    &\lesssim \lambda^{s+d + \max\{n-d-x,0\} + \max\{x-s-1,0\}+\epsilon} \leq \lambda^{\max\{d+s, n-1, \frac{n-1}{2}+s\}+\epsilon},
\end{align*}
where the last inequality follows from minimizing the function $x \mapsto \max\{n-x,d\}+\max\{x-1,s\}$ in the interval $0\leq x\leq \frac{n+1}{2}$. Thus since $d \leq n-2$ for any $\sigma \not \in \{\sigma_{\ell}\}_{0\leq \ell \leq m}$, estimate \eqref{eq:edgerest} follows in this case.

For the second case, we first assume that $F_k \cap F_j = \emptyset$. Under this assumption, the faces $F_j$ and $F_k$ of the re-scaled polytope $\Omega_\lambda$ are a distance $\sim \lambda$ away of each other. So close to $F_j$, respectively $F_k$, we have $h_n(|r-\sigma_k r|)\lesssim \lambda^{-\frac{n+1}{2}}$, respectively $h_n(r-\sigma_j r|) \lesssim \lambda^{-\frac{n+1}{2}}$. Thus again from Lemma \ref{lem:reflectionbound},
\begin{align*}
    \lambda^s E_{\sigma_j,\sigma_k}(\lambda) \lesssim \lambda^s \int_{\Omega_\lambda^2} \lambda^{-\frac{n+1}{2}}(1+|r-r'|)^{-\frac{n+1}{2}}|r-r'|^{-s} \,\mathrm{d}r \,\mathrm{d}r' \lesssim \lambda^{\max\{n-1,\frac{n-1}{2}+s\}+\epsilon}.
\end{align*}
Finally, if $F_j \cap F_k \neq \emptyset$, then the normal vectors $n_j, n_k$ are not parallel. Consequently, the variables $w_j = \pi_{\sigma_j}^\perp r+\pi_{\sigma_j}^\perp r' \in \R$, $w_k=\pi_{\sigma_k}^\perp r + \pi_{\sigma_k}^\perp r' \in \R$ and $r-r' \in \R^n$ are independent. Therefore, we can split the decay of $h_n$ and use Lemma~\ref{lem:reflectionbound} to compensate for the integration in the directions $w_j,w_k$ and $r-r'$. This yields the estimate
\begin{align*}
    \lambda^s E_{\sigma_k,\sigma_j}(\lambda) &\lesssim \lambda^s \int_{\substack{|r-r'|\lesssim \lambda \\ |w_j|+|w_k|\lesssim \lambda}}\frac{(1+|w_j|)^{-1} (1+|\pi_{\sigma_j}(r-r')|)^{-\frac{n-1}{2}}}{(1+|w_k|)(1+|\pi_{\sigma_k}(r-r')|)^{\frac{n-1}{2}}}|r-r'|^{-s}\,\mathrm{d}r\,\mathrm{d}r' \\
    &\lesssim \lambda^{\max\{n-2+s+,n-1\}+\epsilon},
\end{align*}
which completes the proof of the lemma. \end{proof}

From Lemma \ref{lem:lowerorder} and the symmetric relation $E_{\sigma,\tau}(\lambda) = E_{\tau,\sigma}(\lambda)$, we see that only the terms $E_{\sigma_\ell,\sigma_\ell}(\lambda)$ and $E_{\sigma_0,\sigma_\ell}(\lambda)$ (where $\sigma_0$ is the identity in $\R^n$) gives significant contributions. We thus need to compute their asymptotics. Let us start with the term $E_{\sigma_0,\sigma_0}(\lambda)$. In this case, from Lemma \ref{lem:overlap},  the decay of $h_n$, and the change of variables $z = r-r'$, we find that
\begin{align}     &E_{\sigma_0,\sigma_0}(\lambda) \nonumber \\
&=  \int_{\Omega_\lambda-\Omega_\lambda}\frac{|h_n(|z|)^2}{|z|^s} \int_{(\Omega_\lambda-z)\cap \Omega_\lambda)}\,\mathrm{d} r'\,\mathrm{d} z \nonumber \\
&=\int_{\Omega_\lambda-\Omega_\lambda} \frac{h_n(|z|)^2}{|z|^s}\biggr(\lambda^n |\Omega| - \lambda^{n-1}\int_{\partial \Omega} (z\scpr n(r'))_+ \,\mathrm{d}\mathcal{H}^{n-1}(r') + \lambda^{n-2} \mathcal{O}(|z|^2)\biggr)\,\mathrm{d}z \nonumber \\
&= \lambda^n|\Omega| \int_{\R^n} \frac{h_n(|z|)^2}{|z|^s}\,\mathrm{d}z - \lambda^{n-1}|\partial \Omega|\int_{\R^{n-1}\times [0,\infty)}\frac{h_n(|z|)^2 z_n}{|z|^s} \,\mathrm{d}z + \mathcal{O}(\lambda^{n-2+\max\{1-s,0\}}). \label{eq:00term}
\end{align}
Next, let us look to the terms $E_{\sigma_j,\sigma_j}$ with $j\geq 1$. For simplicity, let us assume without loss of generality that $F_j \subset \R^{n-1}\times \{0\}$ and $n_j = (0,...,0,1)$. Let us also denote the height of $\Omega$ by $H= \max\{r_n : \R^{n-1}\times \{r_n\} \cap \Omega \neq \emptyset \}$, the cross-section of $\Omega$ at height $h$ by $\Omega(h) = \{ r \in \R^{n-1} : (r,h) \in \Omega\}$, and the projection sending $(r_1,...,r_n) \in \R^n$ to $(r_1,...,r_{n-1}) \in \R^{n-1}$ by $\pi_n$. Since $\Omega$ is a convex polytope, we can bound the area of the symmetric difference of the cross-sections at different heights by $|\Omega(h)\triangle \Omega(h')|\lesssim |h-h'|$. In particular, a scaling argument yields
\begin{align}
    \bigr|\bigr(\Omega_{2\lambda}(h+z_n)-\pi_n z\bigr) \cap \bigr(\Omega_{2\lambda}(h-z_n)+\pi_nz\bigr)\bigr| - |\Omega_{2\lambda}(0)| \lesssim \lambda^{n-2} (|h|+|z|), \label{eq:crosssectionest}
\end{align}
for any $z \in \Omega_\lambda$. We can now use the above estimate with the change of variables $z = r-r',w = r+r'$ and the decay of $h_n$ to obtain
\begin{align}
    &E_{\sigma_j,\sigma_j}(\lambda) \nonumber \\
    &= \int_{\Omega_\lambda-\Omega_\lambda} \int_{(\Omega_{2\lambda}-z)\cap (\Omega_{2\lambda}+z)} \frac{h_n(|(\pi_nz,w_n)|)^2}{|z|^s}\frac{\,\mathrm{d}z\,\mathrm{d}w}{2^n} \nonumber \\
    &= \int_{\Omega_\lambda-\Omega_\lambda}\int_{|z_n|}^{2\lambda H-|z_n|} \frac{h_n(|(\pi_n z,w_n)|)^2}{|z|^s} |\Omega_{2\lambda}(0)|\frac{\,\mathrm{d} w_n\,\mathrm{d}z}{2^n} + \mathcal{O}(\lambda^{\max\{n-1-s,n-2\}+\epsilon}) \nonumber \\
    &= \frac{\lambda^{n-1}|F_j|}{2} \int_{\R^n} \int_{|z_n|}^\infty \frac{h_n(|(\pi_n z,w_n)|)^2}{|z|^s} \,\mathrm{d} w_n \,\mathrm{d} z+ \mathcal{O}(\lambda^{\max\{n-1-s,n-2\}+\epsilon}), \label{eq:jjterm}
\end{align}
where we used that $\,\mathrm{d}z\,\mathrm{d}w = 2^n \,\mathrm{d}r\,\mathrm{d}r'$, $|\Omega_{2\lambda}(0)| = 2^{n-1}\lambda^{n-1} |F_j|$, and that
\begin{align*}
    &\lambda^{n-2} \int_{|z|\lesssim \lambda} \int_{|z_n|}^{2H\lambda} (1+|(\pi_n z, w_n)|)^{-n-1}|z|^{-s}(|w_n|+|z|)\,\mathrm{d}w_n\,\mathrm{d}z \lesssim \lambda^{\max\{n-1-s,n-2\}+\epsilon},\\
    &\lambda^{n-1}\int_{|z|\lesssim \lambda} \int_{2H\lambda-|z_n|}^{\infty} (1+|(\pi_n z, w_n)|)^{-n-1}|z|^{-s}\,\mathrm{d}w_n\,\mathrm{d}z \lesssim \lambda^{\max\{n-2,n-1-s\}+\epsilon}, \quad \mbox{and} \\
    &\lambda^{n-1}\int_{|z| \gtrsim \lambda} \int_{|z_n|}^{\infty}(1+|(\pi_n z, w_n)|)^{-n-1}|z|^{-s}\,\mathrm{d}w_n\,\mathrm{d}z \lesssim \lambda^{n-1-s+\epsilon}
\end{align*}
for any $\epsilon>0$. For the last terms, $E_{\sigma_0,\sigma_j}(\lambda)$ with $1\leq j \leq m$, one can use the same change of coordinates together with \eqref{eq:crosssectionest} to find that
\begin{multline}
    E_{\sigma_0,\sigma_j}(\lambda) = \frac{\lambda^{n-1}|F_j|}{2}\int_{\R^n}\int_{|z_n|}^\infty\frac{h_n(|(\pi_n z, w_n)|)h_n(|z|)}{|z|^s}\,\mathrm{d}w_n \,\mathrm{d}z \\ + \mathcal{O}(\lambda^{\max\{n-1-s,n-2\}+\epsilon}). \label{eq:0jterm}
\end{multline}
Hence by summing \eqref{eq:00term},\eqref{eq:jjterm}, and \eqref{eq:0jterm} with the estimates in Lemma \ref{lem:lowerorder} we conclude that
\begin{multline}
    E_x^{\textnormal{ctm}}(\lambda) = c_{x,1}(n,s) \lambda^{n+s} + \bigr(c_{FS}(n,s) + c_{BL}(n,s)\bigr)\lambda^{n-1+s} \\ + \mathcal{O}(\lambda^{\max\{n-1,n-2+s,\frac{n-1}{2}+s\}+\epsilon}), \label{eq:summaryest}
\end{multline}
with the constants $c_{x,1}, c_{FS}$ and $c_{BL}$ defined according to Theorem~\ref{exchangethm}. 

Finally, to complete the proof we just need to bound the difference $E^{\textnormal{ctm}}_x(\lambda) - E_x(\lambda)$. For this, we use Corollary \ref{Lpthm}, Theorem \ref{offdiagonalthm}, and the decay of $h$ (Lemma \ref{lem:fouriertransformsphere}) to obtain the estimate
\begin{align}
    \int_{\Omega\times\Omega}\frac{|S_\lambda|^2}{|r-r'|^s}-&\frac{|S^{\textnormal{ctm}}_\lambda|^2}{|r-r'|^s} \,\mathrm{d}x \,\mathrm{d}y \notag \\
    &\leq \int_{\Omega \times \Omega} \frac{|S_\lambda - S^{\textnormal{ctm}}_\lambda|^2}{|r-r'|^s} + \frac{2|S_\lambda - S^{\textnormal{ctm}}_\lambda||S^{\textnormal{ctm}}_\lambda|}{|r-r'|^s} \,\mathrm{d}r\,\mathrm{d}r' \nonumber \\
    &\lesssim \norm{S_\lambda - S^{\textnormal{ctm}}_\lambda}_{L^p}^2 \norm{|r-r'|^{-s}}_{L^q(\Omega^2)} + \norm{S_\lambda - S^{\textnormal{ctm}}_\lambda}_{L^\infty}\int \frac{|S^{\textnormal{ctm}}_\lambda|}{|r-r'|^s} \nonumber  \\
    &\lesssim \lambda^{2n-2-\frac{n-1}{n+1}\left(2+\frac{2(n-1)}{p}\right)}\norm{|r|^{-s}}_{L^q(\Omega_2)} + \lambda^{(n-1) \frac{n}{n+1} +\max\{\frac{n-1}{2},s\}}\log \lambda \label{eq:referenceest}
\end{align}
where $\frac{2}{p}+\frac{1}{q} = 1$ and the $\log \lambda$ term is just needed for the case $\frac{n-1}{2} = s$ (e.g. Coulomb in 3D). Now given $\epsilon>0$ we can choose $q< n/s$ such that $2/p = 1- s/n - \epsilon$. For such $q$, the function $|r|^{-s}$ belongs to $L^q_{\mathrm{loc}}(\R^n)$ and the first term in \eqref{eq:referenceest} is of order $\lambda^{n-1 + s(n-1)^2/(n^2+1)+ \epsilon}$. Therefore, 
\begin{align}
    E_x(\lambda) = E^{\textnormal{ctm}}_x(\lambda) + \mathcal{O}\biggr(\lambda^{n-1 +s \frac{(n-1)^2}{n^2+n} + \epsilon} + \lambda^{(n-1) \left(\frac{3}{2} - \frac{1}{n+1}\right)} \log \lambda \biggr), \label{eq:ctmexchange}
\end{align}
which together with \eqref{eq:summaryest} completes the proof of Theorem \ref{exchangethm} for the Dirichlet case. For the Neumann case, one just need to change the sign before the terms $E_{\sigma_\ell,\sigma_0}(\lambda)$. For the periodic case, one replaces $E_{\sigma,\tau}(\lambda)$ by
\begin{align*}
	E_{v,w}(\lambda) = \int_{\Omega_\lambda^2} \frac{h_n(|r-r'+\lambda v|)h_n(|r-r'+\lambda w|)}{|r-r'|^s}\,\mathrm{d}r\,\mathrm{d}r',
\end{align*}
where $v,w \in \mathcal{R}_\Omega^{nb}$. By using arguments similar to the ones presented above, one can show that all the terms $E_{v,w}(\lambda)$ with $v \neq 0$ or $w\neq 0$ give lower order contributions. The proof then reduces to computing the asymptotic expansion of $E_{0,0}(\lambda) = E_{\sigma_0,\sigma_0}(\lambda)$, which we already did (see \eqref{eq:00term}).

\appendix

\section{Strictly tessellating polytopes}\label{sec:tesselatingpolytopes}

We now show that our definition of a strictly tessellating polytope is equivalent to \cite[Definition 2]{rowlett2021crystallographic}. 

\begin{proposition} Let $\Omega \subset \R^n$ be an open polytope. Then $\Omega$ is strictly tessellating in the sense of Definition~\ref{def:kaleidoscopic} if and only if $\R^n = \bigcup_{j \in \N} \overline{\Omega_j}$, where each $\Omega_j$ is obtained by reflecting $\Omega$ across its boundary faces and the hyperplanes extending the boundary faces of each $\Omega_j$ have empty intersection with (the interior of) $\Omega_k$ for any $j, k \in \N$.
\end{proposition}

\begin{proof} First, let us assume that $\Omega$ is strictly tessellating in the sense of \cite[Definition 2]{rowlett2021crystallographic} and then show that $\Omega$ satisfies Definition~\ref{def:kaleidoscopic}. For this, first observe that by \cite[Corollary 1]{rowlett2021crystallographic}, all eigenfunctions of the Dirichlet Laplacian $-\Delta_\Omega$ are trigonometric, thus real analytic in $\R^n$. Lam\'e's fundamental theorem (see \cite[Theorem 4]{rowlett2021crystallographic}) then implies that any eigenfunction $e_j$ is anti-symmetric with respect to reflection over the faces of $\Omega$, and therefore, $e_j(r) = \det \sigma e_j(\sigma r)$ for any $\sigma \in \mathcal{R}_\Omega$. Now suppose that $\sigma(\Omega) = \tau(\Omega)$ for some $\sigma, \tau \in \mathcal{R}_\Omega$. Then we have $(\tau^{-1} \sigma)(\Omega) = \Omega$ and $\det (\tau^{-1}\circ\sigma) e_j(\tau^{-1} \sigma r) = e_j(r)$ for any eigenfunction $j \in \N$. But since $\{e_j\}_{j \in \N}$ is an orthonormal basis of $L^2(\Omega)$, the push-back map $f \mapsto \det (\sigma \circ \tau) \bigr( \sigma^\# \circ (\tau^{-1})^\# \bigr)f$ is the identity in $L^2(\Omega)$, which shows that $\tau = \sigma$ and $\Omega$ satisfies Definition~\ref{def:kaleidoscopic}.

For the converse implication, just note that $\Omega$ clearly tessellates $\R^n$ with reflected copies of itself, hence, it is enough to show that the hyperplanes extending the boundary faces of any reflected polytope do not intersect the interior of $\Omega$. So let $\sigma \in \mathcal{R}_\Omega$ and $H_\ell$ be the hyperplane extending the face $\sigma(F_\ell)$ of $\sigma(\Omega)$. Then the reflection over $H_\ell$ is given by the composition $\tau_\ell = \sigma \circ \sigma_\ell \circ \sigma^{-1}\in \mathcal{R}_\Omega$ where $\sigma_\ell$ is the reflection over the face $F_\ell$ of $\Omega$. As a consequence, if we suppose that $H_\ell \cap \Omega \neq \emptyset$, then we have $\tau_\ell (\Omega) \cap \Omega \neq \emptyset$ because $H_\ell$ is invariant under the reflection $\tau_\ell$. But from our definition of strictly tessellating polytopes, this implies that $\tau_\ell$ is the identity, which contradicts the fact that $\tau_\ell$ is a reflection over the hyperplane $H_\ell$. We thus conclude that $H_\ell \cap \Omega = \emptyset$, which completes the proof.
\end{proof}
Next, we prove the characterization of the intersection $\overline{\Omega}\cap \overline{\sigma(\Omega)}$ that was used in the proof of Lemma~\ref{lem:reflectionbound}.
\begin{lemma}[Intersection characterization]\label{lem:faces} Let $\Omega = \{ r \in \R^n : r\scpr n_j < \alpha_j, 1\leq j \leq m\}$ be a strictly tessellating polytope with faces $F_\ell = \{r \in \overline{\Omega} : r \scpr n_\ell = \alpha_\ell \}$. Suppose that $I_\sigma = \overline{\Omega} \cap \overline{\sigma(\Omega)} \neq \emptyset$ for some $\sigma \in \mathcal{R}_\Omega^{nb}\setminus \{\sigma_0\}$. Then there exists $j_1,...,j_p$ such that $I_\sigma = \bigcap_{k=1}^p F_{j_k}$, $\sigma \in \langle \sigma_{j_1},...,\sigma_{j_p}\rangle$, and the interior
\begin{align}
	{\textrm int} \bigcap_{k=1}^p F_{j_k} = \biggr\{ r \in \R^n : r\scpr n_{j} \begin{dcases} = \alpha_j \quad &\mbox{if $j = j_k$ for some $1\leq k \leq p$.} \\
	< \alpha_j &\mbox{otherwise.} \end{dcases} \biggr\}
\end{align}
is non-empty. Here (and in the proof below) $\langle \sigma_{j_1},...,\sigma_{j_p} \rangle$ denotes the group generated by $\sigma_{j_1},...,\sigma_{j_p}$.
\end{lemma}
\begin{proof} 
The result follows if we show the following claim:
\begin{align}
	\mbox{{\textbf Claim:} ${\textrm int} \bigcap_{k=1}^p F_{j_k}$ is contained in the interior of $\bigcup_{\sigma \in \langle \sigma_{j_1},...,\sigma_{j_k}\rangle} \overline{\sigma(\Omega)}$.} \label{eq:claim}
\end{align}
Indeed, if this holds, then we can argue as follows. Since $\partial \Omega$ is the union of the interior of all possible face intersections, for any $\tau \in \mathcal{R}_\Omega^{nb}$ we can find $q \in {\textrm int} \bigcap_{k=1}^p F_{j_k} \cap \partial \overline{\tau(\Omega)}$ for some faces $\{F_{j_k}\}_{k\leq p}$. By the claim, the non-empty open set $B_\delta(q) \cap \tau(\Omega)$ is contained in $\bigcup_{\sigma \in \langle \sigma_{j_1},...\sigma_{j_p} \rangle} \overline{\sigma(\Omega)}$ (for $\delta$ small) and must intersect some $\sigma(\Omega)$ (because $\bigcup_{\sigma \in \langle \sigma_{j_1},...,\sigma_{j_k} \rangle} \partial \sigma(\Omega)$ is a countable union of sets with Hausdorff dimension $n-1$). By the strictly tessellating property, we have $\tau = \sigma \in \langle \sigma_{j_1},...,\sigma_{j_p}\rangle$, hence $\bigcap_{k=1}^p F_{j_k} \subset \partial \tau(\Omega)$. Moreover, if $j_1,...,j_p$ is minimal in the sense that ${\textrm int} \bigcap_{\ell =1}^{p-1} F_{j_{k_\ell}} \cap \partial \tau(\Omega) = \emptyset$ for all possible choice $\{j_{k_\ell}\}_{\ell=1}^{p-1} \subset \{j_k\}_{k=1}^p$, then $\overline{\tau(\Omega)}\cap \overline{\Omega} = \bigcap_{k=1}^p F_{j_k}$ (by a convexity argument), which proves the lemma.

To prove the claim we use induction and argue by contradiction. First, it is clear that ${\textrm int} F_k$ is contained in the interior of $\overline{\Omega} \cap \overline{\sigma_k(\Omega)}$. Now suppose that the claim holds for intersections of $n-1$ faces and there exists some $q \in  {\textrm int} \bigcap_{k=1}^n F_{j_k}$ for which 
\begin{align*}
	B_\delta(q) \not\subset C_n \coloneqq \bigcup_{\sigma \in \langle \sigma_{j_1},...,\sigma_{j_n}\rangle} \overline{\sigma(\Omega)} \quad \mbox{ for any $\delta>0$ small.}
\end{align*} 
Since $B_\delta(q)\setminus \bigcap_{k=1}^n F_{j_k}$ is open and connected (as $n \geq 2$) and $C_n$ is closed and has non empty interior inside $B_\delta(q)$, there exists some point $q_2 \in  B_\delta(q) \cap \partial C_n \setminus \bigcap_{k=1}^n F_{j_k}$. In particular, $q_2 \in \partial \sigma(\Omega)$ for some $\sigma \in \langle \sigma_{j_1},...\sigma_{j_n} \rangle$. Moreover, since $\sigma$ is an isometry that leaves the intersection $\bigcap_{k=1}^n F_{j_k}$ invariant, we have $\sigma^{-1}(q_2) \in B_\delta(q) \cap \partial \Omega \setminus \bigcap_{k=1}^n F_{j_k}$. Therefore, if $\delta>0$ is small enough, $\sigma^{-1}(q_2)$ must be contained in the interior of the intersection of at most $n-1$ of the faces $\{F_{j_k}\}_{k\leq n}$.  By assumption, this implies that $\sigma^{-1}(q_2)$ is in the interior of $C_n$. But since $\sigma$ is an isometry (hence open) and $C_n$ is invariant under $\sigma$, we conclude that $q_2$ belongs to the interior of $C_n$, contradicting the fact that $q_2 \in \partial C_n$.

\end{proof}

\section{The wave kernel method}\label{sec:wavekernel}

In this section, we recall some well-known results about the homogeneous wave equation (see, e.g., \cite{sogge2014hangzhou,safarov1997asymptotic} for more detailed discussions). These classical results are used in Section~\ref{sec:wavekernelpolytope} to construct the exact wave kernel on strictly tessellating polytopes and fundamental domains of lattices, which is the key ingredient in the proofs of Theorems~\ref{offdiagonalthm} and ~\ref{L2thm}.

We start by recalling some classical existence, uniqueness and regularity results for the solutions of the wave equation on bounded domains. In what follows, we assume $\Omega \subset \R^n$ to be an open, bounded, and connected domain with Lipschitz boundary. Then, let us consider the initial value problem (IVP) for the wave equation in $\Omega$,
\begin{align}
    \begin{dcases} \partial_{tt}u - \Delta u = 0  \quad &\mbox{ in } \Omega \times \R, \\
    \partial_t u(r,0) = 0, \\
    u(r,0) = g(r)  &\mbox{for some $g \in C^\infty_c(\Omega)$, }\end{dcases} \label{eq:waveequationomega}
\end{align}
with the boundary conditions (BCs)
\begin{align}
    \begin{dcases} u(r,t) = 0  \quad &\mbox{ on $\partial \Omega \times \R$ (Dirichlet BCs), or} \\
    \nabla_r u(r,t)\scpr n(r) = 0 \quad &\mbox{ on $\partial \Omega \times \R$ (Neumann BCs),} \end{dcases} \label{eq:boundaryconditions}
\end{align}
where $n(r)$ is the unit normal vector to $\partial \Omega$ at $r$ and $v\scpr w = \sum_{j=1}^n v_j w_j$ is the standard scalar product in $\R^n$. Then, for an initial condition $g \in C^\infty_c(\Omega)$, the unique solution to \eqref{eq:waveequationomega}\eqref{eq:boundaryconditions} in $C^\infty(\Omega\times \R)$ is given by
\begin{align*}
    u(r,t) = \bigr(\cos(t\sqrt{-\Delta_\Omega})g\bigr)(r),
\end{align*}
where $\Delta_\Omega$ is the self-adjoint extension of the Laplacian in $\Omega$ defined by the boundary conditions, and $\cos(t\sqrt{-\Delta_\Omega})$ is defined via the spectral calculus. (We refer the reader to \cite[Chapter 6]{taylor1996partial} for a proof.) In particular, if $u$ is the solution of \eqref{eq:waveequationomega} for some $g\in C^\infty_c(\Omega)$, then from the spectral theorem we have
\begin{align}
    \int_{\R} f(t) u(r,t) \,\mathrm{d} t = \bigr(\widehat{f}(\sqrt{-\Delta_\Omega})g\bigr)(r) \label{eq:prewavekernelid}
\end{align}
for any $f\in S(\R)$ even (i.e. $f(s) = f(-s)$ for any $s\in\R$). The identity above lies at the heart of the wave equation method in spectral asymptotics because it allows us to obtain information on the kernel of $\widehat{f}(\sqrt{-\Delta_\Omega})$ through (approximate) solutions of \eqref{eq:waveequationomega}. 

\begin{remark*} If $\Omega$ is the fundamental domain of a lattice, then periodic boundary conditions can be imposed and the same results described above hold.
\end{remark*}

To construct the wave kernel on bounded domains, we will need an explicit representation of the wave kernel in $\R^n$ and its finite speed of propagation property. For later use, we state it as a lemma here.
\begin{lemma}[Wave kernel on $\R^n$ \cite{sogge2014hangzhou}] \label{lem:wavekernelRn} Let $E_0(t)$ be the distribution defined by
\begin{align}
    \inner{E_0(t),g}_{\mathcal{D}'(\R^n),\mathcal{D}(\R^n)} = \frac{1}{(2\pi)^n}\int_{\R^n} \cos(t|k|) \widehat{g}(k) \,\mathrm{d} k, \quad \mbox{for $g \in C_c^\infty(\R^n)$.} \label{eq:E0def}
\end{align}
Then, $E_0(t) \in \mathcal{E}^\prime(\R^n)$ (where $\mathcal{E}^\prime$ is the set of distributions with compact support) and $\mathrm{supp}(E_0(t)) = \{r\in \R^n : |r|\leq |t|\}$. Moreover, for any $g\in C^\infty(\R^n)$, the function defined by
\[ u(r,t) \coloneqq \bigr(E_0(t) \ast g\bigr)(r) = \frac{1}{(2\pi)^n} \int_{\R^n} \cos(t|k|)\widehat{g}(k) e^{i k \scpr r} \,\mathrm{d} k\]
is smooth and satisfies the wave equation in $\R^n\times \R$ with initial condition $u(r,0) = g$ and $\partial_t u(r,0) = 0$.
\end{lemma}

\section*{Acknowledgment}
 I am grateful to Gero Friesecke for suggesting the problem and encouraging feedback on the results. I also want to thank Victor Ivrii for a brief but helpful email exchange regarding the contents of Theorems~\ref{offdiagonalthm} and \ref{L2thm}, and Mi-Song Dupuy for his suggestions on the initial draft of this paper.

\bibliographystyle{abbrv}
\bibliography{main}

\end{document}